\documentclass{aa}  
\usepackage{graphicx}
\usepackage{txfonts}
\def\degrees{$^{\circ}$}
\def\srm{$\sigma_{\rm RM}~$}
\def\rmm{$\langle{\rm RM}\rangle~$}

\begin{document}
   \title{Structure of the magnetoionic medium around the FR Class I radio galaxy 3C\,449}
   \subtitle{}

 \author{D. Guidetti\inst{1,2,3}
         \and
         R. A. Laing\inst{1}
         \and 
         M. Murgia\inst{4}
         \and
         F. Govoni\inst{4}
         \and
         L. Gregorini\inst{3}
         \and
         P. Parma\inst{2}
         }

 \offprints{dguidett@eso.org}

   \institute{European Southern Observatory, Karl-Schwarzschild-Stra$\beta$e 2, D-85748 Garching-bei-M\"unchen, Germany
              \and
              INAF - Istituto di Radioastronomia,
              Via Gobetti 101, I--40129 Bologna, Italy
              \and
              Dipartimento di Astronomia, Univ. Bologna,
              Via Ranzani 1, I--40127 Bologna, Italy 
              \and
              INAF - Osservatorio Astronomico di Cagliari,
              Loc. Poggio dei Pini, Strada 54, I--09012 Capoterra (CA), Italy
               }

   \date{Received; accepted}

  \abstract
  {}
{The goal of this work is to constrain the strength and structure of the magnetic field 
associated with the environment of the  radio source 3C\,449, using 
observations of Faraday rotation, which we model with a structure function 
technique and by comparison with numerical simulations.  We assume that 
the magnetic field is a Gaussian, isotropic random variable and that it 
is embedded in the hot intra-group plasma surrounding the radio source.}
 {For this purpose, we present detailed rotation measure images for the polarized radio source 3C\,449, 
previously observed with the Very Large Array at seven frequencies between 1.365 and 8.385\,GHz.
All of the observations are consistent with pure foreground Faraday rotation. We quantify the 
statistics of the magnetic-field fluctuations by deriving rotation 
measure structure functions, which we fit using models derived from 
theoretical power spectra. We quantify the errors due to sampling by 
making multiple two-dimensional realizations of the best-fitting power 
spectrum. We also use depolarization measurements to estimate the minimum 
scale of the field variations. We then make three-dimensional models with 
a gas density distribution derived from X-ray observations and a random 
magnetic field with this power spectrum. By comparing our simulations with 
the observed Faraday rotation images, we can determine the strength of the 
magnetic field and its dependence on density, as well as the outer scale 
of magnetic turbulence.}
{Both rotation measure and depolarization data are consistent with a broken power-law magnetic-field power spectrum,
with a break at about 11\,kpc and slopes of 2.98 and 2.07 at smaller and larger scales respectively.
The maximum and minimum scales of the fluctuations are  $\approx$65 and $\approx$0.2\,kpc,
respectively. The average magnetic field strength at the cluster centre is 3.5$\pm$1.2\,$\mu$G, 
decreasing linearly with the gas density within $\approx$16\,kpc of the nucleus. At larger 
distances, the dependence of field on density appears to flatten, but this 
may be an effect of errors in the density model. The magnetic field is not 
energetically important.}
{}


   \maketitle
  
%

\section{Introduction}
\label{sec:intro}
Magnetic fields in the hot plasma associated
with groups and clusters of galaxies are poorly understood, but are thought to
play a vital role in regulating thermal conduction (e.g.\ Balbus 2000; Bogdanovic et
al. 2009) and influencing the dynamics of 
cavities formed by radio jets (e.g.\ Dursi \& Pfrommer 2008; O'Neill et al.\ 2009).
The existence of magnetic fields can be demonstrated in several different ways 
(e.g.\ Carilli \& Taylor 2002; Govoni \& Feretti 2004 
and references therein).
One of the ways of studying these fields is
via the  Faraday effect: rotation of the plane of linearly polarized radiation by
a magnetized plasma. Synchrotron emission from radio sources (either behind or
embedded within the group/cluster medium) can be used to probe the distribution
of foreground Faraday rotation. These can be combined with X-ray
observations (which provide the thermal gas density profile) to infer the strength
and fluctuation properties of the magnetic field.

Faraday rotation studies of clusters have been carried out using both
statistical samples of background radio sources (e.g. Lawler \& Dennison 1982, Vall\'ee et al. 1986,
Kim et al. 1990, Kim et al. 1991, Clarke et al. 2001)  or individual radio
sources within the clusters
(e.g. Taylor \& Perley 1993; Feretti et al. 1995; Feretti et al. 1999a,b; Govoni et
al. 2001; Eilek \& Owen 2002; Pollack et al. 2005; Govoni et al. 2006; Guidetti
et al. 2008). 
The central magnetic field strengths deduced from these data are usually a few
$\mu$G, but can exceed 10\,$\mu$G in the inner regions 
of relaxed cool-core clusters (e.g.\ Taylor et al. 2002). 
The RM distributions of radio galaxies in both interacting and relaxed clusters
are generally patchy, indicating that cluster magnetic
fields show structure on scales $\la 10$\,kpc.

Several studies of Abell clusters (Murgia et al. 2004; Govoni et al. 2006;
Guidetti et al. 2008) have shown that detailed RM images of radio galaxies can
be used to infer not only the strength of the cluster magnetic field, but also
its power spectrum.  The analysis of Vogt \& En{\ss}lin (2003, 2005) suggests
that the power spectrum has a power law form with the slope appropriate for
Kolmogorov turbulence and that the auto-correlation length of the magnetic field
fluctuations is a few kpc.  The deduction of a Kolmogorov slope could be
premature, however: there is a degeneracy between the slope and the outer
scale which is difficult to resolve with current Faraday-rotation data (Murgia
et al. 2004; Guidetti et al. 2008; Laing et al. 2008). Indeed, Murgia et
al. (2004) pointed out that shallower magnetic field power spectra are possible if
the magnetic field fluctuations have structure on scales of several tens of kpc.  Recently,
Guidetti et al. (2008) showed that a power-law power spectrum with a Kolmogorov
slope and an abrupt long-wavelength cut-off at 35\,kpc gave a very good fit to
their Faraday rotation and depolarization data for the radio galaxies in A2382,
although a shallower slope extending to longer wavelengths was not ruled out.

While most work until recently has been devoted to rich clusters of galaxies,
little attention has been given in the literature to sparser environments,
although similar physical processes are likely to be at work.  Faraday-rotation
fluctuations have previously been detected in galaxy groups (e.g. Perley et al.,
1984; Feretti et al. 1999a), but without deriving in
detail the geometry and structure of the magnetic field.  The first detailed
work on galaxy groups was done by Laing et al. (2008), who analysed the radio
emission of 3C\,31. 

They found that the three-dimensional magnetic-field power spectrum
$\widehat{w}(f)$,defined in Sect.~\ref{2d-general}, might be
described in terms of spatial frequency $f$ by a broken power law
$\widehat{w}(f)\propto D_0 f^{~-q}$ with $q=11/3$\footnote{$q=11/3$
is the slope of the three-dimensional power spectrum for Kolmogorov turbulence.} for
$f>$0.062\,arcsec $^{-1}$ (corresponding to a spatial scale of about 17\,kpc) and q=2.32 at lower frequencies, although a
power spectrum with a slope of 2.39 and an abrupt cut-off at high frequencies could
not be ruled out. Their results are qualitatively similar to those for sources
in Abell clusters.

Magnetic fields associated with galaxy groups deserve to be
investigated in more detail, since their environments are more representative than 
those of rich clusters. Moreover, observations, analytical models and MHD simulations
of galaxy clusters all suggest 
that the magnetic-field intensity should scale with the thermal
gas density 
(e.g. Brunetti et al. 2001, Dolag 2006, Guidetti et al. 2008). A key
question is whether the  relation between magnetic field strength and density in
galaxy groups is a continuation of this trend.\\

This paper presents a detailed analysis of Faraday rotation
in 3C\,449, a bright, extended radio source hosted by the central galaxy
of a nearby group. With the aim of shedding new light on the environment
around this source, we derive the statistical properties of the magnetic field
from observations of Faraday rotation, following the method developed by Murgia et al. (2004).
We use numerical simulations to predict the Faraday rotation for different
strengths and power spectra of the magnetic field.

The paper is organized as follows.
In Sect.\,\ref{general}  the general properties of the radio source under investigation are presented.
Sect.\,\ref{vla} presents the radio images
on which our analysis is based.  In Sect.\,\ref{sec:rm_obs}, we discuss the
observed Faraday rotation distribution of 3C\,449 and assess the contribution
from our Galaxy. The observed depolarization and its relation to
the RM properties are investigated in Sect.\,\ref{sec:dp}. Our two-dimensional
analysis of the structure of the RM fluctuations and a three-dimensional model of the magnetic
field consistent with these results are presented in 
Sect.\,\ref{2d} and Sect.\,\ref{sec:model}, respectively.  Sect.\,\ref{sec:sum} summarizes our conclusions and briefly compares the Faraday-rotation properties
of 3C\,449 with those of other sources.

Throughout this paper we assume a cosmology with $H_0$ = 71 km
s$^{-1}$Mpc$^{-1}$, $\Omega_m$ = 0.3, and $\Omega_{\Lambda}$ = 0.7, which
implies that 1\,arcsec corresponds to 0.342\,kpc at the distance of 3C\,449.

\section{The radio source 3C\,449: general properties}
\label{general}

In this paper,  we image and model the Faraday rotation distribution across the
giant Fanaroff-Riley Class I (FR\,I; Fanaroff \& Riley 1974) radio source
3C\,449, whose environment is very similar to that of 3C\,31.  
The optical counterpart of 3C\,449, UGC\,12064, is a dumb-bell galaxy  and is the most prominent member of the group of galaxies 2231.2+3732 (Zwicky \& Kowal 1968).
The source is relatively nearby (z=0.017085, RC3.9, De Vaucouleurs et al. 1991)
and quite extended, both in angular (30\arcmin) and linear size, so it is an
ideal target for an  analysis of the Faraday rotation distribution: detailed
images can be constructed that can serve as the basis of an accurate study of
magnetic field power spectra.

\begin{figure*}
\centering
\includegraphics[width=13cm]{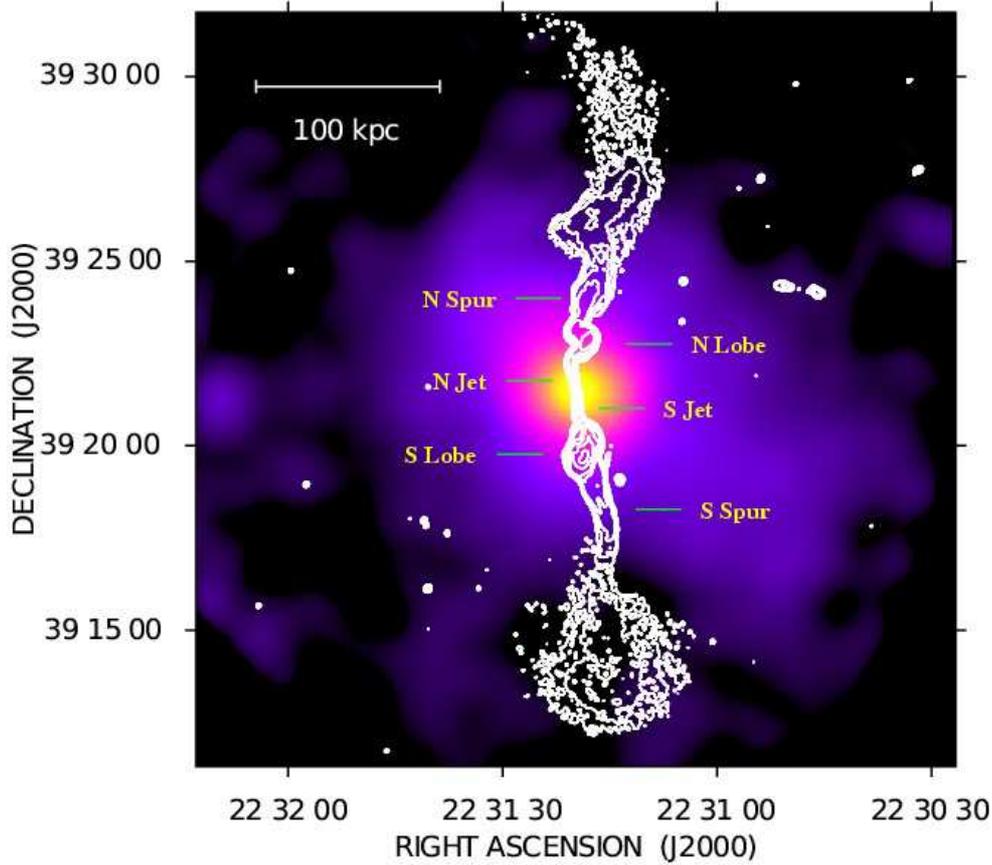}
\caption[]{Radio contours of 3C\,449 at 1.365\,GHz superposed on the XMM-Newton X-ray image (courtesy of J. Croston,
Croston et al. 2003).
The radio contours start at 3$\sigma_{I}$ and increase by factors of 2. The restoring beam is 5.5\,arcsec FWHM.
The main regions of 3C\,449 discussed in the text are labelled.\label{XMM}}
\end{figure*}

3C\,449 was one of the first radio galaxies studied in detail with the VLA (Perley et. al 1979).
High- and low-resolution radio data already exist 
and the source has been mapped at many frequencies.
The radio emission of 3C\,449 (Fig.~\ref{XMM}) is elongated in the N--S direction and is characterized by long,
two-sided \textit{jets} with striking mirror symmetry close to the nucleus. The jets
terminate in well-defined inner \textit{lobes}, which fade 
into well polarized \textit{spurs}, of which the southern one is more
collimated. The spurs in turn expand to form diffuse outer lobes.

The brightness ratio of the radio jets  is very nearly 1, implying that they 
are close to the plane of the sky if they are intrinsically symmetrical and have 
relativistic flow velocities similar to those derived for other FR\,I jets (Perley et
al. 1979, Feretti et al. 1999a, Laing \& Bridle 2002).
In this paper, we therefore assume that the jets lie exactly in the plane of the sky,
simplifying the geometry of the Faraday-rotating medium.

Hot gas associated with the galaxy has been detected on both the group and galactic scales by X-ray imaging (Hardcastle et al. 1998; Croston et al. 2003).
These observations revealed deficits in the X-ray  surface brightness
at the positions of the outer radio lobes, suggesting interactions with the surrounding
material. Fig.\,\ref{XMM} shows radio contours at 1.365\,GHz overlaid on the X-ray
  emission as observed by the XMM-Newton satellite (Croston et al. 2003).
The X-ray radial surface brightness profile
of 3C\,449 derived from these data can be fitted  with the sum of a point-source convolved with
the instrumental response and a 
$\beta$ model (Cavaliere \& Fusco-Femiano, 1976),
\begin{equation} 
\label{beta}
n_e(r) = n_0 (1+r^2/r_c^2)^{-\frac{3}{2}\beta}\,,
\end{equation}
where $r$, $r_c$ and $n_0$ are the 
distance from the group  X-ray centre, the group core radius and the central electron density, respectively.
Croston et al.\ (2008) found a best fitting model with $\beta = 0.42 \pm 0.05$,
$r_c$ = 57.1\,arcsec and $n_0 =3.7 \times 10^{-3}$cm$^{-3}$.
In what follows, we assume that the group gas density is described 
by the model of Croston et al. (2008).  The X-ray depressions noted by Croston et
al.\ (2003) are at distances larger than those at which we can measure linear polarization, and
there is no direct evidence of smaller cavities close to the nucleus. We
therefore neglect any departures from the spherically-symmetrical density model,
noting that this approximation may become increasingly inaccurate where the
source widens (i.e.\ in the inner lobes and spurs).

3C\,449 resembles 3C\,31 in environment and in radio morphology: both sources
are associated with the central members of groups of galaxies and their
redshifts are very similar.
The nearest neighbours are at a projected distances of about 30\,kpc
in both cases.  Both radio sources have large angular extents, bending jets and
long, narrow tails with low surface brightnesses and steep spectra, although
3C\,31 appears much more distorted on large scales.  There is one significant
difference: the inner jets of 3C\,31 are thought to be inclined by
$\approx$50$^\circ$ to the line of sight (Laing \& Bridle 2002), whereas those
in 3C\,449 are likely to be close to the plane of the sky (Feretti et al.\
1999a). We therefore expect that the magnetized foreground medium will be very  
similar in the two sources, but that the geometry will be significantly
different, leading to a much more symmetrical distribution of Faraday rotation
in 3C\,449 compared with that observed in 3C\,31 by Laing et al.\ (2008).

\section{Total intensity and polarization properties}
\label{vla}

The Very Large Array (VLA\footnote{The Very Large Array is a facility of the National Science Foundation, operated under 
cooperative agreement by Associated Universities, Inc.}) observations and their reduction have been presented  by  Feretti et al. (1999a). 
The high quality  of these data make this source suited for a very detailed analysis  of the statistics of the Faraday rotation.

We produced total intensity ($I$) and polarization ($Q$ and $U$) images at
frequencies in the range 1.365 -- 8.385\,GHz from the combined, self-calibrated
u-v datasets described by Feretti et al.\ (1999a).  The centre frequencies and
bandwidths are listed in Table~\ref{pol}.  Each frequency channel was imaged
separately, except for those at 8.245 and 8.445\,GHz, which were averaged. The
analysis below confirms that these frequency-bandwidth combinations lead to
negligible Faraday rotation across the channels, as already noted by Feretti et
al. (1999a).  All of the datasets were imaged with Gaussian tapering in the u-v
plane to give resolutions of 1.25\,arcsec and 5.5\,arcsec FWHM, {\cal CLEAN}ed
and restored with circular Gaussian beams.  The first angular resolution is the
highest possible at all frequencies and provides good signal-to-noise for the
radio emission within 150\,arcsec ($\simeq$50\,kpc) of the radio core (the well
defined radio jets and the inner lobes), while minimizing beam
depolarization.  The lower resolution of 5.5\,arcsec allows imaging of the
extended emission as far as 300\,arcsec ($\simeq$100\,kpc) from the core at
frequencies from 1.365 -- 4.985\,GHz (the 8.385-GHz dataset does not have
adequate sensitivity to image the outer parts of the source).  We can therefore
study the structure of the magnetic field in the spur regions, which lie well
outside the bulk of the X-ray emitting gas. Noise levels for both sets of
images are given in Table~\ref{pol}.  Note that the maximum scales of structure
which can be imaged reliably with the VLA at 8.4 and 5\,GHz are $\approx$180 and
$\approx$300\,arcsec, respectively (Ulvestad, Perley \& Chandler 2009).  For
this reason, we only use the Stokes $I$ images for quantitative analysis
within half these distances of the core.  The $Q$ and $U$ images have much less
structure on such large scales and are reliable to distances of $\pm$150\,arcsec
at 8.4\,GHz and $\pm$300\,arcsec at 5\,GHz,  limited by sensitivity rather than
systematic errors due to missing flux as in the case of $I$ image.

Images of polarized intensity $P = (Q^2+U^2)^{1/2}$ (corrected for Ricean bias, following Wardle \&
Kronberg 1974),
fractional polarization $p=P/I$ and polarization angle $\Psi=(1/2)\arctan(U/Q)$ were derived from the $I$, $Q$, and $U$ images.
\begin{table*}
\caption{Parameters of the total intensity and polarization images. Col. 1:
Observation frequency. Col. 2: Bandwidth (note that the images at 8.385\,GHz are
derived from the average of two frequency channels, both with bandwidths of
50\,MHz, centred on 8.285 and 8.485\,GHz). Cols.3, 4 : rms noise levels in total
intensity ($\sigma_{I}$) and linear polarization ($\sigma_{QU}$, the average of
$\sigma_{Q}$ and $\sigma_{U}$) at 1.25\,arcsec FWHM resolution; Col. 5: mean
degree of polarization at 1.25\,arcsec; Col. 6, 7: rms noise levels for the
5.5\,arcsec images; Col. 8: mean degree of polarization at 5.5\,arcsec.  We
estimate that the uncertainty in the degree of polarization, which is dominated
by systematic deconvolution errors on the $I$ images, is $\approx$0.02 at each
frequency.\label{pol}} \centering
\begin{tabular} {c c c c c c c c}  
\hline\hline
$\nu$    & Bandwidth & \multicolumn{3}{c} {1.25 arcsec} & \multicolumn{3}{c} {5.5 arcsec} \\  
         &      & $\sigma_{I}$ & $\sigma_{QU}$ & $\langle p \rangle$ & $\sigma_{I}$ & $\sigma_{QU}$ & $\langle p \rangle$ \\ 
 (GHz)   &   (MHz)   &  (mJy/beam)   & (mJy/beam)  & & (mJy/beam)   & (mJy/beam) &  \\  
\hline    
&&&&&&&\\  
1.365 & 12.5 & 0.037  & 0.030 & 0.24 & 0.018 & 0.014  & 0.26 \\
1.445 & 12.5 & 0.021  & 0.020 & 0.25 & 0.020  & 0.011  & 0.27  \\
1.465 & 12.5 & 0.048  & 0.049 & 0.25 & 0.019 & 0.013  & 0.25 \\
1.485 & 12.5 & 0.035  & 0.027 & 0.21 & 0.014 & 0.010  & 0.26 \\
4.685 & 50.0 & 0.017  & 0.017 & 0.32 & 0.017 & 0.013  & 0.37    \\
4.985 & 50.0 & 0.018  & 0.017 & 0.33 & 0.017 & 0.016  & 0.39  \\
8.385 & 100.0 & 0.014 & 0.011 & 0.31 & 0.015 & 0.013  & $-$    \\
&&&&&&&\\  
\hline
\end{tabular}
\end{table*}
All of the polarization images (P, $p$, $\Psi$) at a given frequency were blanked
where the rms error in $\Psi >$ 10$^\circ$ at any frequency.  We then calculated the scalar mean 
degree of polarization $\langle p \rangle$ for each frequency and resolution; 
the results are listed in Table~\ref{pol}.
The values of $\langle p \rangle$ are higher at 5.5\,arcsec resolution than at 1.25\,arcsec 
because of the contribution of the extended and highly polarized emission which is not seen at the higher resolution.
At 1.25\,arcsec, where the beam depolarization is minimized, the mean fractional polarization shows a steady increase from 1.365 to 4.685\,GHz, where it reaches an average value of 0.32 and
then remains roughly constant at higher frequencies, suggesting that the
depolarization between 4.685 and 8.385\,GHz is insignificant.

\section{The Faraday rotation in 3C\,449}
\label{sec:rm_obs}

\subsection{Rotation measure images}
\label{sec:rm_images}

A magnetized, ionized medium rotates the plane of polarization of linearly
polarized radiation passing through it as follows:
\begin{equation}
\label{eq:rm}
\Delta\Psi = \Psi(\lambda) - \Psi_0 = \rm RM~\lambda^2\,,
\end{equation}
where  $\Psi(\lambda)$  is the position angle observed at a wavelength  $\lambda$ and
$\Psi_0$ is the intrinsic position angle.  The rotation measure (RM) is related to the electron 
density ($n_e$), the magnetic field along the line-of-sight ($B_{\parallel}$),
and the path-length ($L$)
through the Faraday-rotating medium according to:
\begin{equation}
\label{equaz}
{\rm RM_{~[rad/m^2]}}=812\int_{0}^{L_{[kpc]}}n_{e~[cm^{-3}]}B_{\parallel~[\mu G]}dl\,.
\end{equation}

Images of rotation measure can be obtained for radio sources by fitting to the polarization
angle as a function of $\lambda^2$, taking into account the well-known problem
of  n$\pi$ ambiguities in the observed $\Psi$, as is done for example by the
{\cal AIPS} task {\cal RM}.
Removal of these ambiguities requires observations at many wavelengths well-spaced
 in $\lambda^{2}$.

\begin{figure*}
\includegraphics[width=18cm]{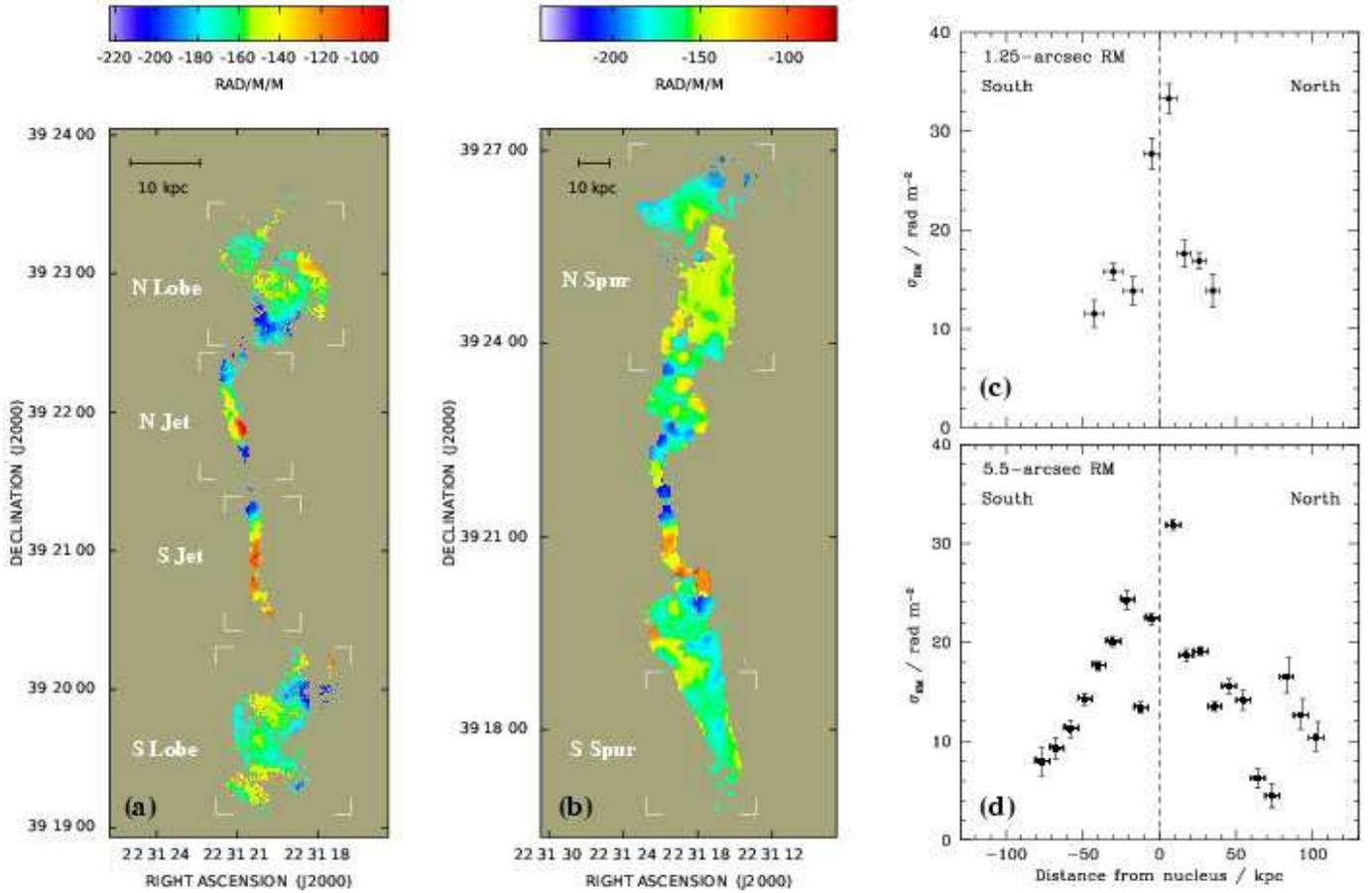}
\caption[]{(a): Image of the rotation measure of 3C\,449 at a resolution of
  1.25\,arcsec FWHM,
computed at the seven
frequencies between 1.365 and 8.385\,GHz.
(b): Image of the rotation measure of 3C\,449 at a resolution of 5.5\,arcsec FWHM,
computed  at the six frequencies between 1.365 and 4.985\,GHz.
In both of the RM images, the sub-regions used for the two-dimensional analysis of Sect.\,\ref{2d} are labelled.
(c) and (d): profiles of \srm as a function of the projected distance from the radio source centre.
The points represent the values of \srm  evaluated in boxes as described in the text.
The horizontal and vertical bars represent the bin widths and the rms on the
mean expected from fitting errors,
respectively. Positive distances are in the direction of the north jet and the vertical dashed lines show the position of the nucleus.\label{highlowrm}
}
\end{figure*}

We produced images of RM and its associated rms error  with resolutions of
1.25\,arcsec  and 5.5\,arcsec (Fig.\,\ref{highlowrm}a and b) using a version of the {\cal AIPS} task {\cal RM}
modified by G. B. Taylor.
The 1.25\,arcsec-RM map was made 
by combining the maps of polarization ${\bf E}$-vector ($\Psi$) at all seven
frequencies available to us, so our sampling of $\lambda^2$ is  very good. 
The RM map was calculated using a weighted least-squares fit
 at pixels with polarization angle uncertainties $<$10\,\degrees at all
frequencies. It is essentially the same as the RM image of Feretti et
al. (1999a), but with more stringent blanking.
The average fitting error is $\simeq$1.4\,rad\,m$^{-2}$  and is  
almost constant over the whole RM image. 
The image of RM at 5.5\,arcsec resolution
was produced using the polarization position angles at the 6 frequencies between
1.365 and 4.985\,GHz (see Table\,\ref{pol}), using the same blanking criterion 
as at higher resolution.

Patches with different size are apparent in the 1.25\,arcsec resolution map,  with
fluctuations down to scales of a few kpc.
The bulk of the RM values range from about $-$220\,rad\,m$^{-2}$ up to
$-$90\,rad\,m$^{-2}$, dominated by the Galactic contribution (see Sect.~\ref{sec:gal}). 
The RM distribution peaks at  $-$161.7\,rad\,m$^{-2}$, with rms dispersion 
\srm=19.7\,rad\,m$^{-2}$. Note that we have not corrected the values of \srm
for the fitting error $\sigma_{\rm RM_{fit}}$. A first order correction would be 
$\sigma_{\rm RM_{true}}=(\sigma_{\rm RM}^2-\sigma^2_{\rm RM_{fit}})^{1/2}$. 
Given the low value for $\sigma_{\rm RM_{fit}}$, the effect of this correction
would be very small.

As was noted by Feretti et al. (1999a), the RM distribution in the inner jets is highly
symmetric about the core with RM $\simeq -197$ rad\,m$^{-2}$ at distances $\la$15\,arcsec. 
The symmetry of the RM distribution in the jets is broken at larger distances from the core: while
the RM structure in the southern jet is homogeneous, with  values around
~$-$130\,rad\,m$^{-2}$, fluctuations  on scales of $\simeq$10\,arcsec ($\simeq$
3\,kpc) around a \rmm of ~$-$160\,rad\,m$^{-2}$ are present in the northern jet.
In both lobes, we observe similar patchy RM structures with
mean values \rmm $\simeq -164$\,rad\,m$^{-2}$ and \srm
$\simeq$16\,rad\,m$^{-2}$.

At 5.5\,arcsec resolution, more extended polarized regions of 3C\,449 can be
mapped with good sampling in $\lambda^2$. The average fitting error is $\simeq$1.0\,rad\,m$^{-2}$.
Both the spurs are characterized by \rmm $\simeq -$160\,rad\,m$^{-2}$, with \srm=15 and 10\,rad\,m$^{-2}$
in  the north  and south, respectively. 
The overall mean and rms for the 5.5\,arcsec image, \rmm $= -$160.7\,rad\,m$^{-2}$ and \srm=18.9\,rad\,m$^{-2}$, 
are very close to those determined at higher resolution and consistent with the
integrated value of $-162 \pm 1$\,rad\,m$^{-2}$ derived by Simard-Normandin et
al. (1981).

\begin{figure}
\includegraphics[height=12.3cm]{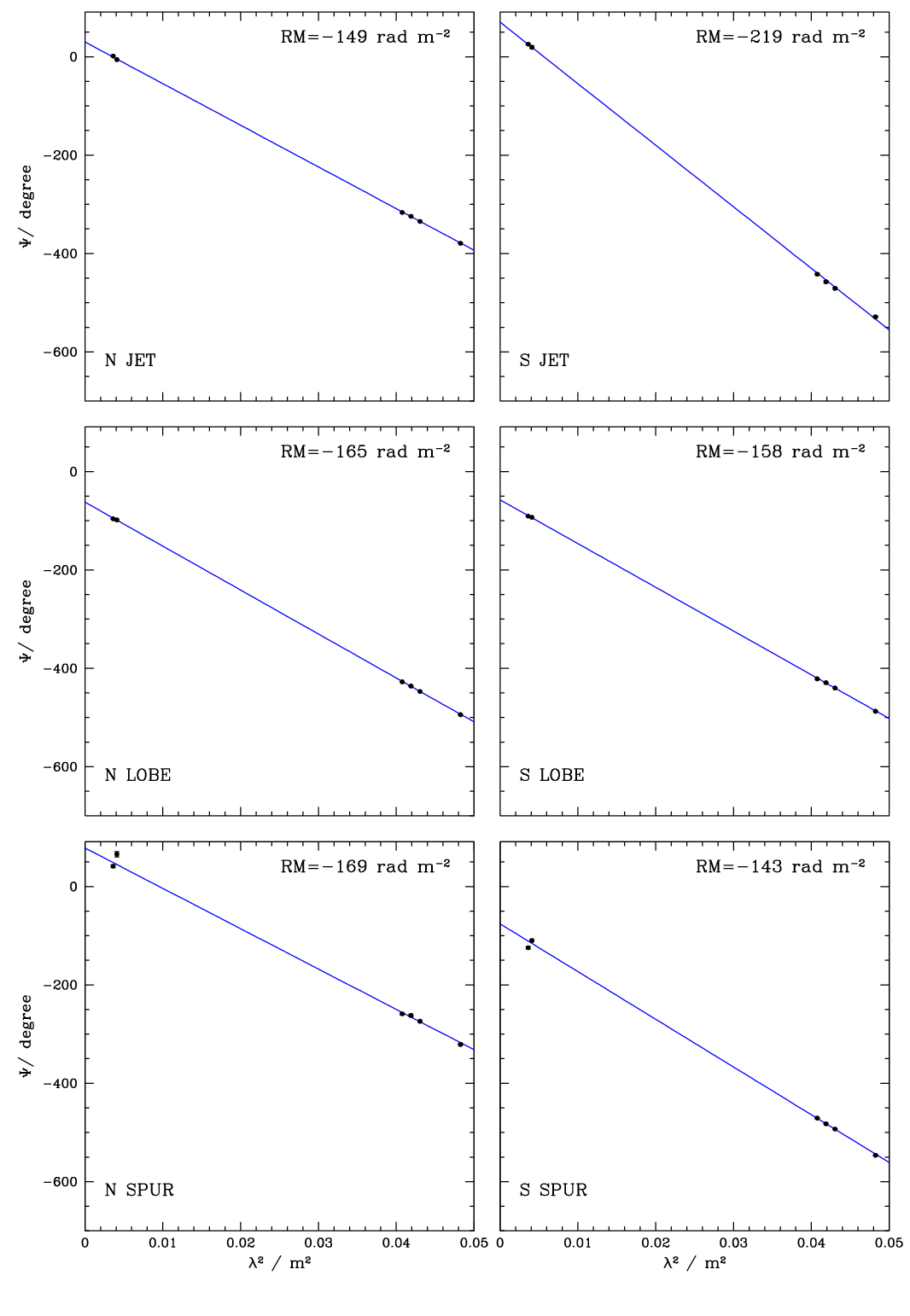}
\caption[]{Plots of  ${\bf E}$-vector position angle $\Psi$ against $\lambda^2$ at representative points of the 5.5-arcsec RM map. 
Fits to the relation $\Psi(\lambda) = \Psi_0 + {\rm RM}\lambda^2$ are shown. The values of
RM are given in the individual panels.\label{fittini}
}
\end{figure}

It was demonstrated by Feretti et al. (1999a) that the polarization position
angles at 1.25\,arcsec resolution accurately follow the relation $\Delta \Psi \propto \lambda^2$ over a
large range of rotation. We find the same
effect at lower resolution: plots of ${\bf E}$-vector position angle
$\Psi$ against $\lambda^2$ at representative points of the 5.5\,arcsec-RM image are shown in Fig\,\ref{fittini}.
As at the higher resolution, there are no significant deviations from 
the relation $\Delta\Psi\propto\lambda^2$ over a range of rotation
 $\Delta\Psi$ of 600\,\degrees, confirming that a foreground magnetized medium is responsible for
the majority of the Faraday rotation and extending this result to regions of
lower surface brightness.

In Fig.\,\ref{highlowrm}(c) and (d), we show profiles of \srm for both low and
high resolution RM images.  The 1.25\,arcsec profile was obtained by averaging
over boxes with lengths ranging from 9 to 13\,kpc along the radio axis; for the
5.5\,arcsec profile we used boxes with a fixed length of 9\,kpc (these sizes were
chosen to give an adequate number of independent points per box).
The boxes extend far enough perpendicular to the source axis to include all unblanked pixels.
In both plots, there is clear evidence for a decrease in the observed \srm towards the
periphery of the source, the value dropping from $\simeq$30\,rad\,m$^{-2}$ close
to the nucleus to $\simeq$10\,rad\,m$^{-2}$ at 50\,kpc. This is qualitatively as
expected for foreground Faraday rotation by a medium whose density (and presumably
also magnetic field strength) decreases with radius.  The symmetry of the \srm
profiles is consistent with our assumption
that the radio source lies in the plane of the sky.

\subsection{The Galactic Faraday rotation}
\label{sec:gal}

For the purpose of this work, 
3C\,449  has an unfortunate line of sight within our Galaxy.
Firstly, the source is located at $l=95.4$\,\degrees~, $b=-15.9$\,\degrees\ in Galactic coordinates, 
where the Galactic magnetic field
is known to be aligned almost along the line of sight.
Secondly, there is evidence from radio and optical imaging for a diffuse,
ionized Galactic feature in front of 3C\,449, perhaps associated with the nearby
HII region S126 (Andernach et al.\ 1992).  
Estimates of the Galactic foreground RM at the position of 3C\,449 from observations of other radio 
sources are uncertain: Andernach et al.\ (1992) found a mean value of
$-$212\,rad\,m$^{-2}$ for six nearby sources, but the spherical harmonic
models of Dineen \& Coles (2005), which are derived by fitting to
the RM values of large numbers of extragalactic sources, predict
$-$135\,rad\,m$^{-2}$.  Nevertheless, it is clear that the bulk of the mean RM
of 3C\,449 must be Galactic.

In order to investigate the magnetized plasma local to 3C\,449, we need to 
constrain the value and possible spatial variation of this Galactic
contribution.  The profiles of \srm (Fig.~\ref{highlowrm}) show that the
small-scale fluctuations of RM drop rapidly with distance from the nucleus. We
might therefore expect the Galactic contribution to dominate on the largest scales. 
At low resolution, we can determine the RM accurately out to $\approx$100\,kpc
from the core. This is roughly 5 core radii for the X-ray emission and
therefore well outside the bulk of the intra-group gas.

\begin{figure}
\includegraphics[height=8cm]{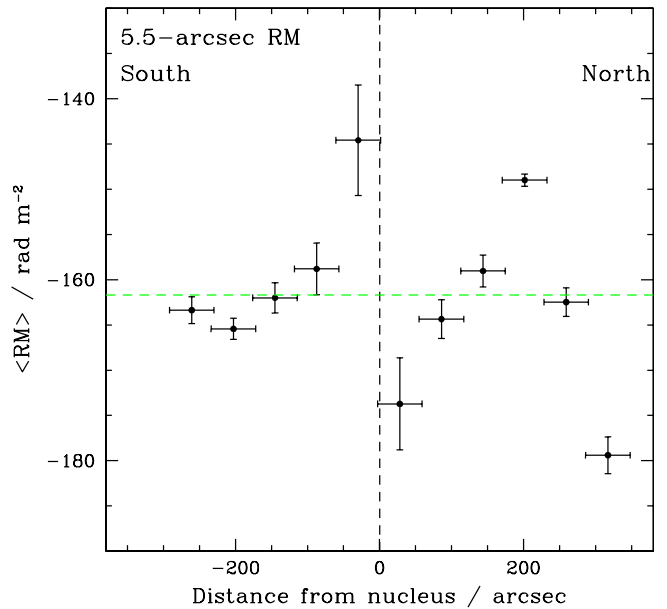}
\caption[]{Profile of RM averaged over boxes of length 20\,kpc along the radio axis for the 5.5\,arcsec image. The horizontal bars represent the bin width. The vertical bars are the errors on the mean calculated from the dispersion in the boxes, the contribution from the fitting error is negligible and is not taken into account.
Positive distances are in the direction of the north jet. The black vertical dashed line indicates the position of the nucleus; the green dashed line shows our adopted mean value for the Galactic RM.\label{low}}
\end{figure}

In order to estimate the Galactic RM contribution, we averaged the 5.5-arcsec RM
image in boxes of length 20\,kpc along the radio axis (the box size has been
increased from that of Fig.~\ref{highlowrm} to improve the display of
large-scale variations).  The profile of \rmm against the distance from the
radio core is shown in Fig.\,\ref{low}. The large deviations from the mean in
the innermost two bins are associated with the maximum in \srm and are almost
certainly due to the intra-group medium. The dispersion in \rmm is quite small
in the south and the value of \rmm = $-$160.7\,rad\,m$^{-2}$ for the whole
source is very close to that of the outer south jet. There are significant
fluctuations in the north, however. Given their rather small scale
($\sim$300\,arcsec), it is most likely that these arise in the local environment
of 3C\,449 and we include them in the statistical analysis given below.

There is some evidence for linear gradients in Galactic RM on arcminute scales:
Laing et al.\ (2006) found a gradient of magnitude
0.025\,rad\,m$^{-2}$\,arcsec$^{-1}$ along the jets of the radio galaxy NGC\,315
($l=124.6$\,\degrees~, $b=-32.5$\,\degrees). They argued that this gradient is
almost certainly Galactic in origin, since the amplitude of the linear variation
exceeds that of the small-scale fluctuations associated with NGC\,315.  In order
to check the effect of a large-scale Galactic RM gradient on our results, we
computed an unweighted least-squares fit of a function \rmm $= {\rm RM}_{0} + ax$,
where $a$ and $\rm RM_{0}$ are constant and $x$ is measured along the radio
axis. The two innermost bins in Fig.\,\ref{low} were excluded from the fit.  Our
best estimate for the gradient is very small:
$a$=0.0054\,rad\,m$^{-2}$\,arcsec$^{-1}$. We have
verified that subtraction of this gradient has a negligible effect on the
structure-function analysis given in Sect.~\ref{sec:sfuncov}.

We therefore adopt a constant value of $-$160.7\,rad\,m$^{-2}$ as the Galactic
contribution.

\section{Depolarization}
\label{sec:dp}

Faraday rotation generally leads to a decrease of the degree of polarization
with increasing wavelength, or \textit{depolarization}.
We define DP$^{\lambda_{1}}_{\lambda_{2}}=p(\lambda_{1})/p(\lambda_{2})$,
where $p(\lambda)$ is the degree of polarization at a given wavelength
$\lambda$.  We adopt the conventional usage in which \textit{higher} depolarization
corresponds to a \textit{lower} value of DP. 

Laing (1984) has summarized the interpretation of polarization data. Faraday
depolarization of radio emission from radio sources can occur in three principal
ways:
\begin{enumerate}
\item  thermal plasma is mixed with the synchrotron
emitting material (\textit{internal depolarization});
\item  there are fluctuations of the foreground 
Faraday rotation
across the beam (\textit{beam depolarization}) and
\item  the polarization angle varies across the finite band of the 
receiving system (\textit{bandwidth depolarization}).
\end{enumerate}

We first estimated the bandwidth effects on the polarized emission of 3C\,449
using the RM measurements from Sect.~\ref{sec:rm_images}. In
the worst case (the highest absolute RM value of $-$240\,rad\,m$^{-2}$) at the
lowest frequency of 1.365\,GHz) the
rotation across the band  is $\approx$10\,\degrees. This results
in a depolarization of 1.7\%, negligible compared with errors due to noise.

If $\lambda^2$ rotation is
observed over a position-angle range $\gg$90\degrees,
then a foreground screen must be responsible for
the bulk of the observed RM.
In that case, depolarization can still 
result from unresolved inhomogeneities of thermal density or magnetic field
in the surrounding medium.
\begin{figure*}
\includegraphics[width=18cm]{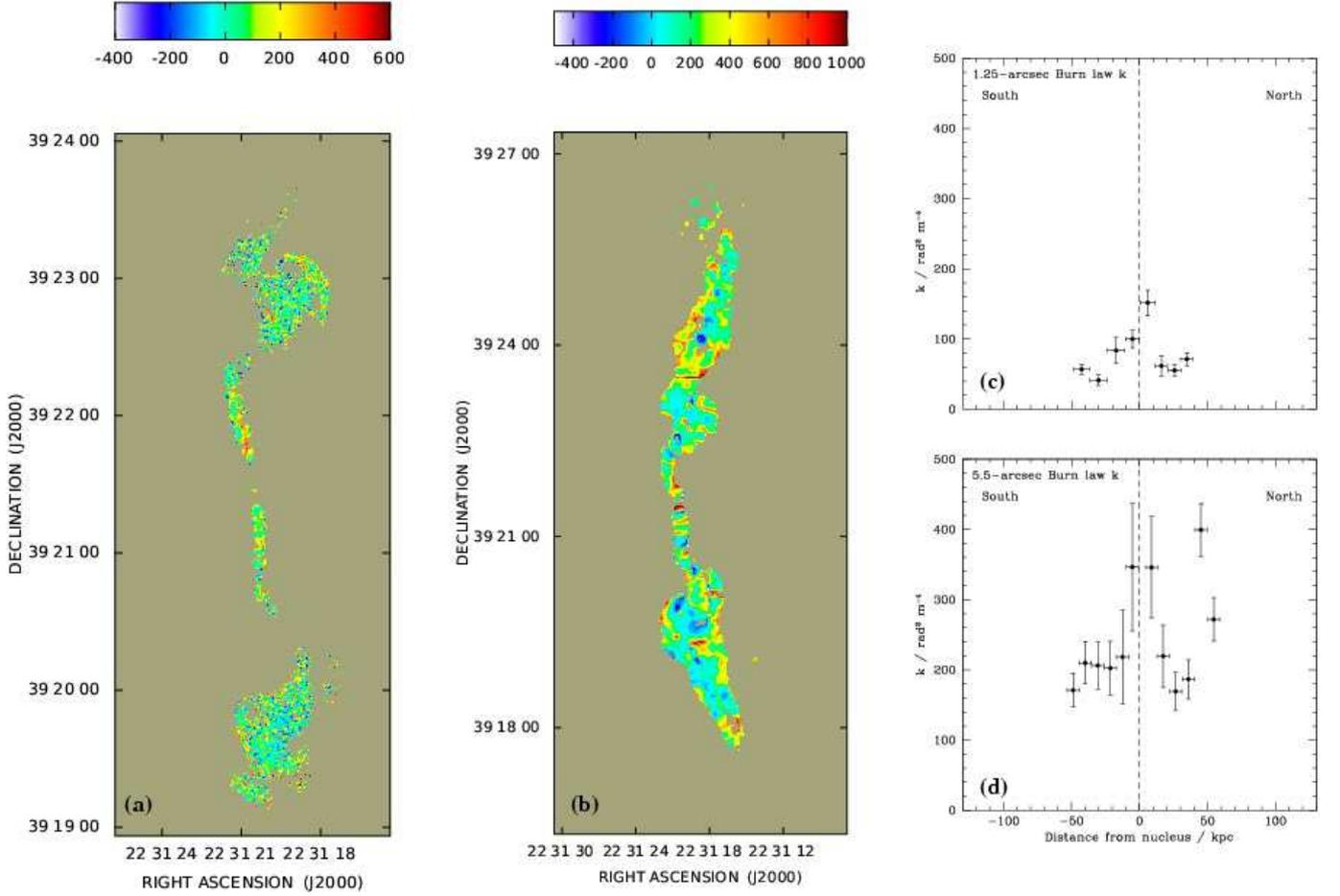}
\caption[]{(a): image of the Burn law $k$ in rad$^2$\,m$^{-4}$
computed from a fit to the relation $p(\lambda)=p(0)\exp(-k\lambda^4)$ for seven frequencies between 1.365 and 8.385\,GHz.
(b): as (a) but the angular resolution is  5.5\,arcsec FWHM, and the $k$ image has been
computed from the fit to the six frequencies between 1.365 and 4.985\,GHz.
(c) and (d): profiles for $k$ as functions of the projected distance from the radio source centre (boxes as in Fig.\,\ref{highlowrm}).
The horizontal and vertical  bars represent  the bin widths  and the error on
the mean, respectively. Positive distances are in the direction of the north jet and the vertical dashed lines show the position of the nucleus.\label{kb}}
\end{figure*}
Our analysis of the depolarization of 3C\,449 is based on the approach of Laing
et al.\ (2008).  In the presence of a foreground Faraday screen with a small
gradient of RM across the beam, it is still possible to observe $\lambda^2$
rotation over a wide range of polarization angle and the wavelength dependence
of the depolarization is expected to follow the Burn law (Burn 1966):
\begin{equation}
\label{equadp}
p(\lambda)=p(0)\exp(-k\lambda^4),
\end{equation}
where $p(0)$ is the intrinsic value of the degree of polarization and
$k$=2$\arrowvert\nabla{RM}\arrowvert^2 \sigma^2$, with ${\rm FWHM} = 2\sigma
(2\ln 2)^{1/2}$.  Since $k \propto\arrowvert\nabla{RM}\arrowvert^2 $, Eq.\,\ref{equadp}
clearly illustrates that higher RM gradients across the beam generate higher $k$
values and hence higher depolarization. The variation of $p$ with wavelength can
potentially be used to estimate fluctuations of RM across the beam which are
below the resolution limit.  We can determine the intrinsic polarization $p(0)$
and the proportionality constant $k$ by a linear fit to the logarithm of the
observed fractional polarization as a function of $\lambda^4$.

 We made images of $k$ at both standard resolutions by weighted least-squares
fitting to the fractional polarization maps, using 
the FARADAY code by M. Murgia. The same frequencies were used as
for the RM images: 8.385 -- 1.365\,GHz and 4.985 --
1.365\,GHz at 1.25 and 5.5\,arcsec resolution, respectively.
By simulating the error distributions for $p$, we established 
that the mean values of $k$ were biased significantly
at low signal-to-noise (cf.\ Laing et al. 2008), so only
data with $p>4\sigma_p$ at each frequency are included in the fits.
We estimate that any bias is negligible compared with the fitting error.
We also derived profiles of $k$ using the same sets of boxes as for
the \srm profiles in Fig.~\ref{highlowrm}.  

\begin{figure}
\includegraphics[height=8.5cm]{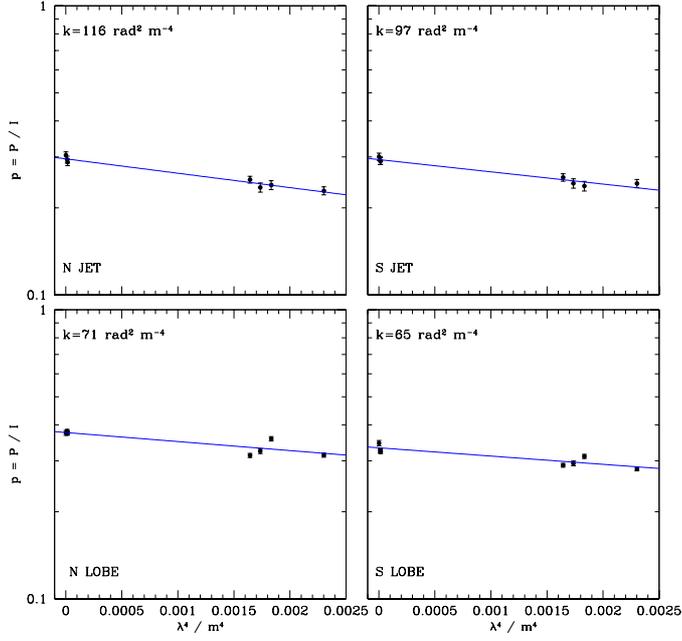}
\caption[]{Plots of degree of polarization, $p$ (log scale) against $\lambda^4$ for representative points at 1.25-arcsec resolution.
Burn law fits (Eq.\,\ref{equadp}) are also plotted. The values of
$k$ are quoted in the individual panels.\label{fittini_dph}}
\end{figure}

The 1.25\,arcsec resolution $k$ map is shown in
Fig.\,\ref{kb}(a), together with the profile of the $k$ values (Fig.\,\ref{kb}c). 
The fit to a $\lambda^4$ law is very good everywhere: examples of
fits at selected pixels in the jets and lobes are shown in in
Fig.\,\ref{fittini_dph}. The symmetry observed in the \srm profiles is also seen in 
the 1.25\,arcsec $k$ image (Fig.\,\ref{kb}): 
the mean values of $k$
are $\simeq$50\,rad$^2$\,m$^{-4}$ for both lobes,  107 and 82\,rad$^2$\,m$^{-4}$ for the 
northern and southern jet, respectively. The region with the highest depolarization 
is in the northern jet,
very close to the core 
and along the west side.
The integrated value of $k$ at this resolution is $\simeq$56\,rad$^2$\,m$^{-4}$,
corresponding to a mean depolarization $DP^{\rm 20 cm}_{\rm 3 cm} \simeq 0.87$.

\begin{figure}
\includegraphics[height=8.5cm]{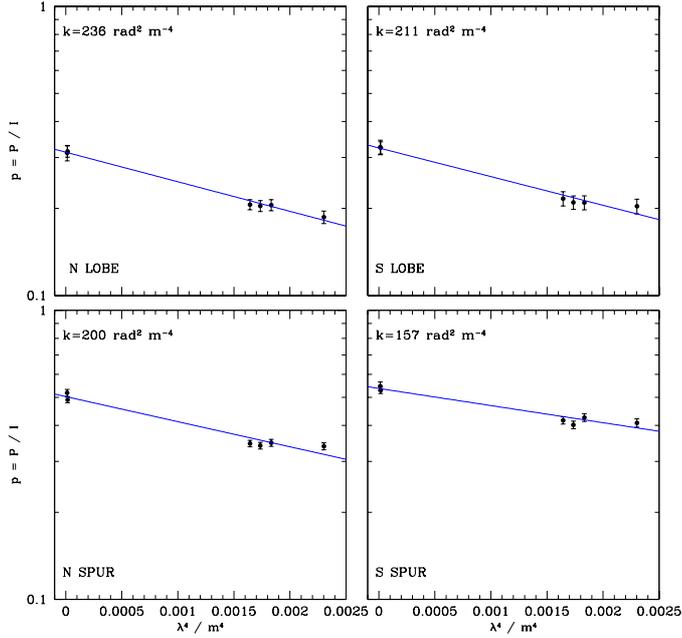}
\caption[]{Plots of degree of polarization, $p$ (log scale) against $\lambda^4$ for representative points at 5.5-arcsec resolution. 
Burn law fits (equation\,\ref{equadp}) are also plotted. The values of
$k$ are quoted in the individual panels.\label{fittini_dpl}}
\end{figure}

The image and profile of $k$ at 5.5-arcsec resolution are shown in Fig.\,\ref{kb}(b) and (d).
The fit to a $\lambda^4$ law is in general 
good and examples are shown in in Figs.\,\ref{fittini_dph}.
As mentioned earlier, 
the maximum scale of structure imaged accurately in
total intensity at 5\,GHz is $\sim$300\,arcsec (100\,kpc) and there are
likely to be significant systematic errors in the degree of polarization on
larger scales. We therefore show the profile only for the inner $\pm$50\,kpc. 
Over this range, the $k$ profiles are quite symmetrical, as at higher resolution.
Note also that the
small regions of very high $k$ at the edge of the northern and southern spurs in
the map shown in Fig.\,\ref{kb} are likely to be spurious.

The mean values of $k$
are $\simeq$184\,rad$^2$\,m$^{-4}$ and 178\,rad$^2$\,m$^{-4}$ for the northern and southern lobes, respectively; 
and $\simeq$238 and 174\,rad$^2$\,m$^{-4}$  in the northern and in the southern spurs.
The integrated value of $k$ is $\simeq$194\,rad$^2$\,m$^{-4}$,
corresponding to a depolarization $DP^{\rm 20 cm}_{\rm 6 cm} \simeq 0.64$.

To summarize, we observe depolarization between 20\,cm and 3\,cm. Since we
measure lower values of $k$ at 1.25\,arcsec than 5.5\,arcsec, there is less
depolarization at high resolution, as expected in the case of beam
depolarization. The highest depolarization is observed in a region of the
northern jet, close to the radio core and associated with a large RM
gradient. Depolarization is significantly higher close to the nucleus,
consistent with the higher path length through the group gas observed in X-rays.
Aside from this global variation, we have found no evidence for detailed correlation of depolarization with 
source structure. Depolarization and RM data are therefore both consistent with a
foreground Faraday screen. We show in Sect.\,\ref{sec:sfunc} that the residual depolarization at 1.25-arcsec resolution
can be produced by RM fluctuations on scales smaller than the beamwidth, but
higher-resolution observations are needed to establish this conclusively.

\section {Two Dimensional Analysis}
\label{2d}

\subsection{General considerations}
\label{2d-general}

In order to interpret the fluctuations of the magnetic field responsible for the
observed RM and depolarization of 3C\,449, we first discuss the statistics of the
RM fluctuations in two dimensions.  We use the notation of Laing et al.\ (2008),
in which ${\bf f}=(f_{x},f_{y},f_{z})$ is a vector in the spatial frequency
domain, corresponding to the position vector ${\bf r} = (x, y, z)$. We take the
$z$-axis to be along the line of sight, so that the vector  ${\bf r}_\perp = (x,
y)$ is in the plane of the sky and ${\bf f}_\perp=(f_{x},f_{y})$ is the
corresponding spatial frequency vector.
Our goal is to estimate the RM power spectrum $\widehat{C}({\bf
f}_{\perp})$, where $\widehat{C}({\bf f}_{\perp})df_{x}df_{y}$ is the power in
the area $df_{x}df_{y}$  and in turn to derive the three dimensional magnetic
field power spectrum $\widehat{w}({\bf f})$, defined so that $\widehat{w}({\bf
f})df_{x}df_{y}df_{z}$ is the power in a volume $df_{x}df_{y}df_{z}$ of
frequency space.

The relation between the magnetic field
statistics and the observed RM distribution is in general quite complicated,
depending on fluctuations in the thermal gas density, the geometry of the source
and the surrounding medium and the effects of incomplete sampling. In order to
derive the magnetic-field power spectrum, we make the following simplifying
assumptions, as in Guidetti et al. (2008) and Laing et al.\ (2008).
\begin{enumerate}
\item The observed Faraday rotation is due entirely to a foreground ionized medium (in
  agreement with our results in Sects\,\ref{sec:rm_obs} and \ref{sec:dp}).
\item The magnetic field is an isotropic, Gaussian random variable, and can
  therefore be characterized by a power spectrum $\widehat{w}(f)$ which is a
  function only of scalar frequency $f$.
\item The form of the magnetic field power spectrum  is independent of  position.
\item The magnetic field is distributed throughout the Faraday-rotating medium, whose density
  is a smooth, spherically symmetric function.
\item The amplitude of $\widehat{w}(f)$ is spatially variable, but is a function
  only of the thermal electron density.
\end{enumerate}
These assumptions guarantee that the spatial distribution of the magnetic field
can be described entirely by its power spectrum $\widehat{w}(f)$, and that for a
medium of constant depth and density, the power spectra of magnetic field and RM
are proportional (En\ss lin and Vogt 2003).

\begin{figure*}
\includegraphics[height=14cm]{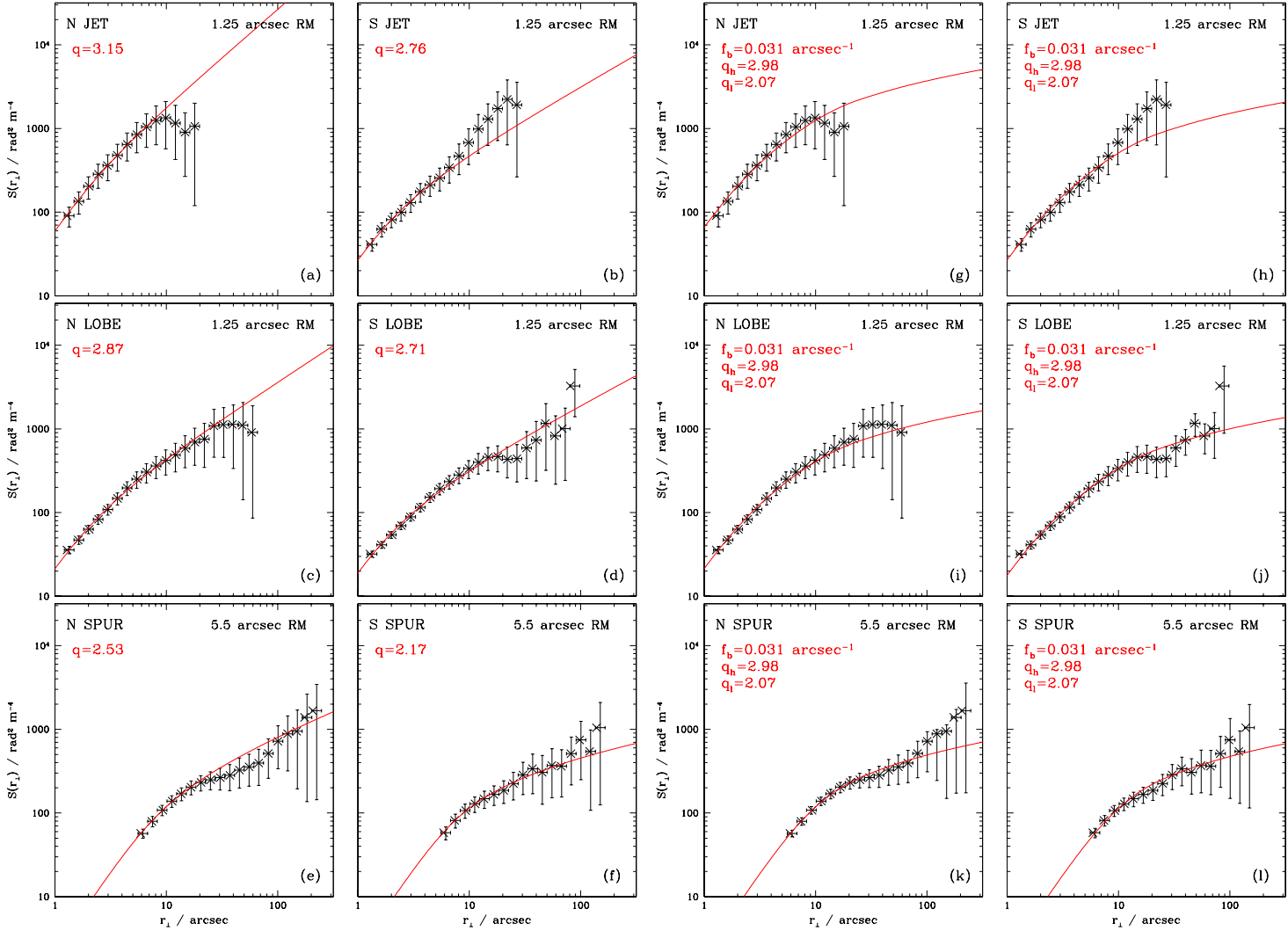}
\caption[]{(a)-(f): Plots of the RM structure functions for the sub-regions showed in Fig.\ref{highlowrm}. The horizontal
bars represent the bin widths and the crosses the centroids for data included in the bins. The red lines are the predictions
for the CPL power spectra described in the text, including the effects of the convolving beam. The vertical error bars
are the rms variations for the structure functions derived using a CPL power
spectrum with the quoted value of $q$ on the observed grid of points for each sub-region.
(g)-(l): as (a)-(f) but using a BPL power spectra with fixed
slopes and break frequency, but variable normalization.
\label{sfunc}}
\end{figure*}

If the fluctuations are isotropic, the RM power spectrum $\widehat{C}(f_{\perp})$  is the Hankel transform of the autocorrelation function  ${C}(r_{\perp})$, defined as
\begin{equation}
 C(r_{\perp})= \langle \rm{RM}({\bf r_{\perp}}+{\bf r'_{\perp}}){\rm RM}({\bf r'_{\perp}}) \rangle, 
\label{auto}
\end{equation}
where ${\bf r_{\perp}}$ and ${\bf r'_{\perp}}$ are vectors in the plane of the sky and $\langle\rangle$ is an average over ${\bf r'_{\perp}}$.

In an ideal case, it would be possible to derive the RM power spectrum and,
consequently, that of the magnetic field, directly from $C(r_{\perp})$.  In
reality, the observations are affected first by the effects of convolution with
the beam, which modify the spatial statistics of RM, and secondly, by the
limited size and 
irregular shape of the sampling region for 3C\,449, which results in a
complicated window function (En{\ss}lin \& Vogt, 2003) and limits the accuracy
with which the zero-level can be determined. In Sect.\,\ref{sec:gal}, we showed
that the 
Galactic contribution to the 3C\,449 RM is large, and we argued that 
a constant value of $-$160.7\,rad\,m$^{-2}$ is the best estimate for its value.
Fluctuations in the Galactic magnetic field on scales comparable with the size of
the radio sources could be present; conversely, the local environment of the
source might make a significant contribution to the mean RM. Both of these
possibilities lead to difficulties in the use of the autocorrelation function.

Laing et al.\,(2008) demonstrated a procedure which takes into account the convolution
effects and minimises the effects of uncertainties in the zero-level.  In
particular, they showed that: 
\begin{enumerate}
\item In the
short-wavelength limit (meaning that changes in Faraday rotation across the beam
are adequately represented as a linear gradient), the measured RM distribution is closely approximated by
the convolution of the true RM distribution with the observing beam.
\item The \textit{structure
function} is a powerful and reliable statistical tool to 
quantify the two dimensional fluctuations of RM, given that it is independent
of the zero level and structure on scales larger than the area under
investigation.
\end{enumerate}
The structure function is defined by
\begin{equation}
S(r_\perp)=\rm{<[RM({\bf r}_\perp + {\bf r}_\perp^\prime)-RM({\bf r}_\perp^\prime)]^2>}  
\label{sfunction}
\end{equation}
(Simonetti, Cordes \& Spangler, 1984; Minter and Spangler 1996). 
It is related to the autocorrelation function $C(r_{\perp})$ for a
sufficiently large averaging region by $S(r_\perp) = 2[C(r_\perp)-C(0)]$.  

Laing et al. (2008) also derived the effects of convolution with the observing
beam on the observed structure function. For the special case of a power-law
power spectrum (their Eq.\ B2), they showed that the observed structure function
after convolution can be heavily modified even at separations up to a few times
the FWHM of the observing beam.  This effect must be taken into account when
comparing observed and predicted structure functions. Laing et al.\ (2008) and
Guidetti et al.\ (2008) also showed that numerical simulation of depolarization
provides complementary information on RM fluctuations on scales smaller than the
beam.

Following the approach of Laing et al. 2008, we initially used the RM structure
function to determine the form of $\widehat{w}(f)$ (Sect.~\ref{sec:sfunc}),
while for its normalization (determined by global variations of density and
magnetic field strength), we made use of three-dimensional simulations
(Sect.\,\ref{sec:model}).

\subsection{Structure functions}
\label{sec:sfunc}

We calculated the structure function for discrete regions of 3C\,449 over which
we expect the spatial variations of thermal gas density, rms magnetic field
strength and path length to be reasonably small. For each of these regions, we
first made unweighted fits of model structure functions derived from power
spectra with simple, parameterized functional forms, accounting for convolution
with the observing beam. We then generated multiple realizations of a Gaussian,
isotropic, random RM field, with the best-fitting power spectrum on the
observed grids, again taking into account the effects of the convolving
beam. Finally, we made a weighted fit using the dispersion of the synthetic
structure functions as estimates of the statistical errors for the \textit{observed}
structure functions, which are impossible to quantify analytically (Laing et
al. 2008).  These errors, which result from incomplete
sampling, are much larger than those due to noise, but depend only weakly on the
precise form of the underlying power spectrum.  Our measure of goodness of fit
is $\chi^2$, summed over a range of separations from $r_\perp =$ FWHM to roughly
half of the size of the region: there is no information in the structure
function for scales smaller than the beam, and the upper limit is set by
sampling.  The errors are, of course, much higher at the large spatial scales,
which are less well sampled.  Note, however, that estimates of the structure
function from neighbouring bins are not statistically independent, so it is not
straightforward to define the effective number of degrees of freedom.

We selected six regions for the structure-function analysis, as shown in
Fig.\,\ref{highlowrm}. These are symmetrically placed about the nucleus,
consistent with the orientation of the radio jets close to the plane of the sky.
For the north and south jets, we derived the structure functions only at
1.25-arcsec resolution, as the low-resolution RM image shows no additional
structure and has poorer sampling.  For the north and south lobes, we computed
the structure functions at both resolutions over identical areas and compared
them. The agreement is very good, and the low-resolution RM images do not sample
significantly larger spatial scales, so we show only the 1.25-arcsec
results. Finally, we used the 5.5-arcsec RM images to compute the structure
functions for the north and south spurs, which are not detected at the higher
resolution.

The structure function has a positive bias given by 2$\sigma_{\rm noise}^2$
where $\sigma_{\rm noise}$ is the uncorrelated random noise in the RM image
(Simonetti, Cordes\& Spangler, 1984).  The mean noise of the 1.25 and 5.5-arcsec RM
maps is $<$1\,rad\,m$^2$ and essentially uncorrelated on scales larger than the
beam.  For each region we therefore subtracted 2$\sigma_{\rm noise}^2$ from the
structure functions, although this correction is always small.  The
noise-corrected structure functions are shown in Fig.\,\ref{sfunc}.

The individual observed structure functions have approximately power-law
forms. Given that the structure function for a power-law power spectrum with no
frequency limits is itself a power law (Minter \& Spangler 1996; Laing et al.\
2008), we first tried to fit the observed data with a RM power spectrum of the
form
\begin{equation}
\label{pure}
\widehat{C}(f_{\perp})\propto~{f_{\perp}^{~-q}}
\end{equation}
over an infinite frequency range. This last assumption allows us to use the 
analytical solution of the structure function, including convolution (Laing et
al. 2008)  and therefore to avoid numerical integration.

The fits were quite good, but systematically gave slightly too much power on
small spatial scales and over-predicted the depolarization. We therefore 
fit a \textit{cut-off power law} (CPL) power spectrum 
\begin{eqnarray}
\widehat{C}(f_{\perp}) &= & 0 ~~~~~~~~~~~~~~~~
f_{\perp}<f_{\rm min}\nonumber \\
                       &= & C_{0}f_{\perp}^{~-q} ~~~~~~~~
f_{\perp}\leq{f_{\rm max}}  \nonumber \\
                       &= & 0 ~~~~~~~~~~~~~~~~
f_{\perp}>f_{\rm max}\,.
\label{cpl}
\end{eqnarray}
Initially, we consider values of $f_{\rm min}$ sufficiently small
that their effects on the structure functions over the observed range of separations
are negligible.
The free parameters of the fit in this case are the slope, $q$, the cut-off
spatial frequency $f_{\rm max}$ and the normalization of the power spectrum,
$C_0$.   
In Table\,\ref{fittingsf}, we give the best-fitting parameters for CPL fits to
all of the individual regions.  
The fitted model structure functions are plotted in 
Fig.\,\ref{sfunc}(a)--(f), together
with error bars derived from multiple realizations of the power spectrum as in
Laing et al.\,(2008).

\begin{table*} 
\caption{CPL power spectrum parameters for the six individual sub-regions of 3C\,449 (lower and upper limits are quoted at $\sim$90\% confidence).
\label{fittingsf}}
\centering       
\begin{tabular}{c c c c c c c c c c c}     
\hline\hline       
 Region     & FWHM & \multicolumn{9}{c} {CPL} \\
            & (arcsec ) & \multicolumn{3}{c} {Best Fit} & \multicolumn{3}{c} {Min Slope} 
	    & \multicolumn{3}{c} {Max Slope} \\
            &    & $q$ &$f_{\rm max}$ & & $q^{\rm -}$ & $f_{\rm max}$ & &
	           $q^{\rm +}$ & $f_{\rm max}$ &  \\
\hline      
N SPUR & 5.50 & 2.53 & 1.96  & & 1.58 & 0.23 & & 3.44 & $\infty$   \\ 
N LOBE & 1.25 & 2.87 & 1.60  & & 2.31 & 0.55 & & 3.35 & $\infty$  \\
N JET  & 1.25 & 3.15 & 1.21  & & 2.29 & 0.30 & & 4.27 & $\infty$  \\
S JET  & 1.25 & 2.76 & 1.95  & & 2.02 & 0.3 & & 3.69 & $\infty$    \\
S LOBE & 1.25 & 2.71 & 1.68  & & 2.36 & 0.65 & & 3.05  & $\infty$   \\
S SPUR & 5.50 & 2.17 & 1.53  & & 0.20 & 0.12  & & 3.95 & 0.12   \\
\hline\hline
\end{tabular}
\end{table*}

In order to constrain RM structure on spatial scales below the beamwidth, we
estimated the depolarization expected from the best power spectrum for each of
the regions with 1.25-arcsec RM images, following the approach of Laing et al.\
(2008).  To do this, we made multiple realizations of RM images on an 8192$^2$
grid with fine spatial sampling. We then derived the $Q$ and $U$ images at our
observing frequencies, convolved to the appropriate resolution and compared the
predicted and observed mean degrees of polarization. 
These values are given in
Table~\ref{fit_burnt}. 
The uncertainties in the expected $<k>$ in Table~\ref{fit_burnt} represent statistical errors determined from multiple realizations of RM images with the same set of power spectrum parameters.
The predicted and observed values are in excellent agreement.
A constant value
of $f_{\rm max} = 1.67$\,arcsec$^{-1}$ predicts very similar values, also listed
in Table~\ref{fit_burnt}.  We have not compared the depolarization data at
5.5-arcsec resolution in the spurs because of limited coverage of large spatial
scales in the $I$ images (Sect.~\ref{vla}), which is likely to introduce
systematic errors at 4.6 and 5.0\,GHz.

We performed a joint fit of the CPL power spectra, minimizing the $\chi^2$ summed over all six
sub-regions, giving equal weight to each and allowing the normalizations to vary
independently. In this case the free parameter of the fit are:
the six normalizations (one for each sub-region) the slope and the maximum spatial frequency.
The joint best-fitting single power-law power spectrum has $q = 2.68$.

A single power law slope does not give a good fit to all of the regions
simultaneously, however.
It is clear
from Fig.\,\ref{sfunc} and Table\,\ref{fittingsf} that there is a flattening in
the slope of the observed structure functions on the largest scales (which are
sampled primarily by the spurs). 
In order to fit all of the data accurately with a single
functional form for the power spectrum, we adopt a 
\textit{broken power law} form (BPL) for the RM power spectrum:
\begin{eqnarray}
\widehat{C}(f_{\perp}) &= & 0  ~~~~~~~~~~~~~~~~~~~~~~~~~~~~~
f_{\perp}<f_{\rm min}\nonumber \\
                       &= & D_{0}f_b^{(q_{\rm l}-q_{\rm h})}f_{\perp}^{~-q_{\rm{l}}} ~~~~~~~~~~~
f_{b}\geq{f_{\perp}}  \nonumber \\
                        &= & D_{0}f_{\perp}^{~-q_{\rm{h}}} ~~~~~~~
f_{\rm max} \geq f_{\perp}  > f_{b} \nonumber \\
                       &= & 0  ~~~~~~~~~~~~~~~~~~~~~~~~~~~~~
f_{\perp} > f_{\rm max}\,.
\label{bpl}
\end{eqnarray}

We performed a BPL joint fit, in the same way as for the CPL power spectra. 
In this case the free parameters of the fit are:
the six normalizations, $D_0$, one for each sub-region, the high and low-frequency slopes, $q_{\rm
h}$ and $q_{\rm l}$, the break and maximum spatial frequency $f_{\rm b}$ and $f_{\rm
max}$. 
We found best fitting parameters of $q_{\rm{l}}$= 2.07,
$q_{\rm{h}}$=2.98, $f_{\rm b}$=0.031\,$\rm{arcsec^{-1}}$. 
As noted earlier, we
also fixed $f_{\rm max} = 1.67$\,arcsec$^{-1}$ to ensure consistency with the
observed depolarizations at 1.25-arcsec resolution.
The corresponding structure functions are plotted in Fig.\,\ref{sfunc}(g)--(l)
and the normalizations for the individual regions are given in
Table~\ref{fit_burnt}.
As for the CPL fits, the errors bars are derived from the rms scatter of the
structure functions of multiple convolved RM realizations.

It is evident from Fig.\,\ref{sfunc}
that the structure functions corresponding to the BPL power spectrum, which gives less power
on large spatial scales, are in much better agreement with the data.
The joint BPL fit  has a  $\chi^2$ of 17.7, compared with 33.5 for the joint CPL fit (the former has only
two extra parameters) confirming this result.

We have so far ignored the effects of any outer scale of magnetic-field
fluctuations. This is justified by the fact that the structure functions for the
spurs continue to rise at the largest observed separations, indicating that the
outer scale must be $\ga$10\,arcsec ($\simeq$30\,kpc). The model structure functions fit to the
observations assume that the outer scale is infinite and the realizations are
generated on sufficiently large grids in Fourier space that the effects of the
implicit outer scale are negligible over the range of scales we sample.  We use
structure-function data for the entire source to determine an approximate value
for the outer scale in Sect.~\ref{sec:sfuncov}.

We now adopt the BPL power spectrum with these parameters and investigate the
spatial variations of the RM fluctuation amplitude using three-dimensional
simulations.

\begin{table*} 
\caption{Best-fitting parameters for the joint CPL and BPL fits to all six sub-regions of 
3C449 (lower and upper limits are quoted at $\sim$90\% confidence).
The values of $q$ and  $f_{\rm max}$  for the joint CPL fit and $q_h$, $q_l$ and  $f_{\rm b}$ for the joint BPL fit are the same for all sub-regions, while the normalizations are varied to minimize the overall $\chi^2$. In the joint BPL fit, the maximum frequency is fixed at $f_{\rm max}=
1.67$\,arcsec$^{-1}$.}
\centering       
\begin{tabular}{c c c c c c c c c c c c c}     
\hline\hline       
&            \multicolumn{5}{c} {Best Fit} & & \multicolumn{3}{c} {Min Slope}& 
	      \multicolumn{3}{c} {Max Slope} \\
&               $q$ &$f_{\rm max}$ &  & $\chi^2$ &  & $q^{\rm -}$  & $f_{\rm max}$     &
	        &   $q^{\rm +}$ &  $f_{\rm max}$ &  \\
&&&&&&&&&&\\
joint CPL  & 2.68 & 1.67 &  &  33.5 &  & 2.55  & 1.30 & &  2.81     & 2.00 &       \\
\hline
  &\multicolumn{5}{c} {Best Fit} & & \multicolumn{3}{c} {Min Slope} & \multicolumn{3}{c} {Max Slope} \\
  & $q_l$ &$f_{\rm b}$ & $q_h$  &  $\chi^2$ & & $q_l^{\rm -}$ & $f_{\rm b}$ &  $q_h^{\rm -}$ &
	  $q_l^{\rm +}$ & $f_{\rm b}$ & $q_l^{\rm +}$  \\
&&&&&&&&&&\\
joint BPL  & 2.07 & 0.031 & 2.97   & 17.7 & & 1.99 & 0.044 & 2.91 & 2.17 & 0.021 & 3.09 \\
&&&&&&&&&&\\
\hline\hline
\label{fittingcomb}
\end{tabular}
\end{table*}

\begin{table*} 
\centering
\caption{Normalizations $C_0$ and $D_0$ for the individual fit parameters corresponding to the CPL, joint CPL and joint BPL fits at a resolution of 1.25\,arcsec. 
Observed and expected depolarizations are also given.
Col.1: region;  Col. 2: observed Burn law $<k>$. Col. 3, 4 and  5: normalization constant $C_0$, fitted maximum spatial frequency $f_{\rm max}$ for the best CPL power spectrum of each region and the predicted $<k>$ for each power spectrum.
Col. 6, 7 as Col. 3 and 5 but for the joint fit to each CPL power spectrum;  Col. 8 and 9 as  Col. 3, and 5 but for
the joint BPL power spectrum. For both the joint CPL and BPL fits, the maximum frequency is fixed at $f_{\rm max}=
1.67$\,arcsec$^{-1}$. In calculating each value of $<k>$ only data with $p>4\sigma_p$ are included.}
\label{fit_burnt}   
\begin{tabular}{c c|c c c|c c|c c c}     
\hline\hline       
 Region       & Observed $<k>$  & \multicolumn{3}{c} {CPL} &  \multicolumn{2}{c} {JOINT CPL} & & \multicolumn{2}{c} {JOINT BPL}  \\
              &                 & $C_0$ & $f_{\rm max}$     & $<k>$ & $C_0$ & $<k>$ &  & $D_0$ & $<k>$  \\
              &  (rad$^2$\,m$^{-4}$)   &  & ($\rm arcsec^{-1}$) &   (rad$^2$\,m$^{-4}$) &    & (rad$^2$\,m$^{-4}$) & &  & (rad$^2$\,m$^{-4}$)   \\ 
\hline      
&&&&&&\\
N LOBE     & 61$\pm$6  & 0.96 & 1.60  &  63$\pm$3 &  1.52   & 66$\pm$3 & &   1.91 & 52$\pm$4\\
N JET      & 106$\pm$12  &  1.34 &   1.21  & 106$\pm$5   & 4.76    & 110$\pm$5   && 0.5 & 109$\pm$5 \\
S JET      & 91$\pm$11  & 1.50  &  1.95 & 70$\pm$5   & 1.94 & 73$\pm$5  & &  1.52 &     65$\pm$4 \\
S LOBE     & 50$\pm$5    & 1.18 &  1.68 &  53$\pm$2   & 1.28   & 50$\pm$2  & &  2.20  &  45$\pm$3 \\
&&&&&&\\
\hline\hline
\end{tabular}
\end{table*}

\section{Three-dimensional analysis}
\label{sec:model}

\subsection{Models}
\label{3Dcode}

We used the software package {\cal FARADAY} (Murgia et al. 2004) to compare the
observed RM with simulated images derived from three-dimensional multi-scale
magnetic-field models.  Given a field model and the density distribution of the
thermal gas, {\cal FARADAY} calculates an RM image by integrating
Eq.\,\ref{equaz} numerically.  As in Sect.~\ref{2d}, we model the fluctuations of RM on the
assumption that the magnetic field responsible for the foreground rotation is an
isotropic, Gaussian random variable and therefore characterized entirely by its
power spectrum.  Each point in a cube in Fourier space is first
assigned components of the magnetic vector potential. The amplitudes are
selected from a Rayleigh distribution of unit variance and the phases are random
in $[0, 2\pi]$.  The amplitudes are then multiplied by the square root of the
power spectrum of the vector potential, which is simply related to that of the
magnetic field.  The corresponding components of the magnetic field along the
line of sight are then calculated and transformed to real space. This procedure
ensures that the magnetic field is divergence-free. The field components in real
space are then multiplied by the model density distribution and integrated along
the line of sight to give a synthetic RM image at the full resolution of the
simulation, which is then convolved to the observing resolution. 

For 3C\,449, we assumed that the source is in a plane perpendicular to the line
of sight which passes through the group centre and simulated the field and
density structure using a 2048$^3$ cube with a real-space pixel size of
0.1\,kpc. We used the best-fitting BPL power spectrum found in
Sect.\,\ref{sec:sfunc}, but with a spatially-variable normalization, as
described below (Sect.~\ref{sec:radial}), and a low-frequency cut-off $f_{\rm
min}$, corresponding to a maximum scale of the magnetic field
fluctuations,\footnote{Here we refer to the scale length $\Lambda$ as a complete
wavelength, i.e. $\Lambda = 1/f$. This differs by a factor of 2 from the
definition in Guidetti et al. 2008, where $\Lambda$ is the reversal scale of the
magnetic field, so $\Lambda = 1/2f$.} $\Lambda_{\rm max}$ ($ = f_{\rm
min}^{-1}$).  The power spectrum of Eq.~\ref{bpl} is then set to 0 for $f <
f_{\rm min}$. We fixed the minimum scale of the fluctuations $\Lambda_{\rm
min} = 0.2$\,kpc. This is equivalent to the value $f_{\rm max} =
1.67$\,arcsec$^{-1}$ found in Sect.\,\ref{sec:sfunc} and also consistent with
the requirement that the minimum scale can be no larger than twice the pixel
size for adequate sampling.

We made multiple synthetic RM images at resolutions of 1.25 and 5.5\,arcsec over
the fields of view of the observations for each combination of parameters.  In
order to estimate the spatial variation of the magnetic-field strength, we first
made a set of simulations with a large, fixed value of $\Lambda_{\rm max}$ and
compared the predicted and observed profiles of \srm
(Sect.~\ref{sec:radial}). We then fixed the radial variation of the field at its
best-fitting form and estimated the value of $\Lambda_{\rm max}$ using a
structure-function analysis for the whole source (Sect.~\ref{sec:sfuncov}).

\begin{table*}
     \caption{Summary of magnetic field power spectrum and density scaling parameters.\label{simul}}
\centering
\begin{tabular}{c c}
\hline
\hline                
\multicolumn{2}{c}{BPL power spectrum}\\
\hline
 $q_{\rm l}=$2.07 &     low-frequency slope \\
 $q_{\rm h}$=2.98 &     high-frequency slope    \\
 $f_{b}$=$\rm 0.031\,arcsec^{-1}$ & break frequency   ($\Lambda_b=1/{f_{b}}$=11\,kpc)\\
 $f_{\rm max}$=$\rm 1.67\,arcsec^{-1}$ & maximum frequency ($\Lambda_{\rm min}=1/{f_{\rm max}}$=0.2\,kpc)\\
 $f_{\rm min}$ fitted  & minimum frequency ($\Lambda_{\rm max}=1/{f_{\rm min}}$) \\
\hline
\multicolumn{2}{c}{Scaling of the magnetic field}\\
\hline
 $B_{0}$ fitted    &    Average magnetic field  at  group centre\\
 $\eta$  fitted    &    Magnetic field exponent of the radial profile: $\langle B\rangle(r)=B_1186
{0}\left[\frac{n_e(r)}{n_0}\right]^{\eta}$  \\     
&\\
\hline
\hline
\label{param}
\end{tabular}
\end{table*}

\subsection{Magnetic field strength and radial profile}
\label{sec:radial}

In order to estimate the radial variation of field strength, we first fixed the
value of the outer scale to be $\Lambda_{\rm max} = 205$\,kpc, the largest
allowed by our simulation grid. Our approach was to make a large number of
simulations for each combination of field strength and radial profile and to
compare the predicted and observed values of \srm evaluated over the boxes used
in Sect.~\ref{sec:rm_images} (Fig.~\ref{highlowrm}).  We used $\chi^2$ summed
over the boxes as a measure of the goodness of fit. This
procedure is independent of the precise value of the outer scale provided that
it is much larger than the averaging boxes. We express our results in
terms of $\chi^2_{\rm
red}$, which is the value of $\chi^2$ divided by the number of degrees of
freedom.

We initially tried a radial field-strength
variation of the form:
\begin{equation}\label{br}
\langle B^2(r)\rangle^{1/2} = B_{0} \left[\frac{n_e(r)}{n_0}\right]^{~\eta}
\end{equation}
as used by Guidetti et al. (2008) and Laing et al. (2008).  Here, $B_{0}$ is the
rms magnetic field strength at the group centre and $n_{e}(r)$ is the thermal
electron gas density, assumed to follow the $\beta$-model profile derived by
Croston et al. (2008; see Sect.\,\ref{general}).
This functional form is consistent with other observations, analytical
  models and numerical simulations. In particular,  $\eta = 2/3$ 
corresponds to flux-freezing and $\eta = 1/2$ to equipartition between thermal
and magnetic energy. Dolag et al. (2001) and Dolag (2006) found $\eta \approx 1$
from the correlation between the observed rms RM and X-ray surface brightness in
galaxy groups and clusters and showed that this is consistent with the results of
MHD simulations.

\begin{figure*}
\centering
\includegraphics[width=11cm]{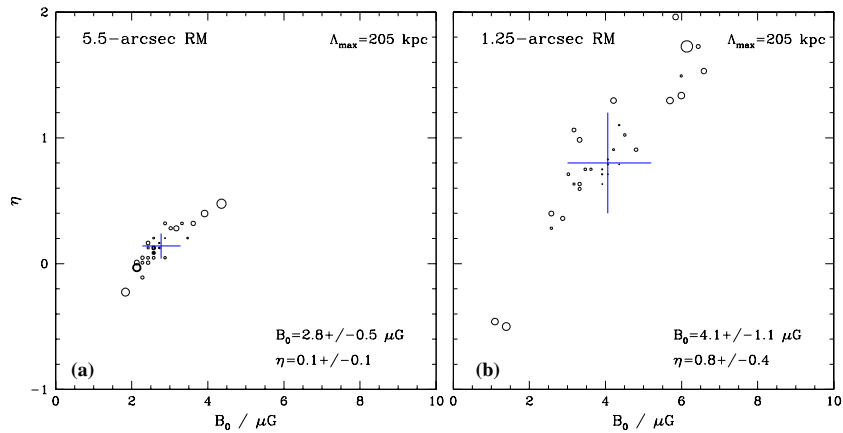}
\caption[]{(a) and (b): Distributions of
  the best-fitting values of $B_{0}$ and $\eta$ from 35 sets of simulations,
  each covering ranges of 0.5 -- 10\,$\mu$G in $B_0$ and 0 -- 2 in $\eta$, at 
  5.5 and 1.25\,arcsec respectively. The sizes of the circles are proportional to
  $\chi^2$ for the fit 
  and the blue crosses represent the means of the distributions
  weighted by $1/\chi^2$.
  The plot shows the expected degeneracy between $B_{0}$ and
  $\eta$. \label{first_cd}}
\end{figure*}

We produced simulated RM images for each combination of $B_0$ and $\eta$ in the ranges
0.5 -- 10\,$\mu$G in steps of 0.1\,$\mu$G  and 0 -- 2 in steps of 0.01, respectively.
We then derived the synthetic \srm profiles and, by comparing them with
the observed one, calculated the unweighted $\chi^2$.
We repeated this procedure 35 times at each angular resolution,
noting the ($B_0$, $\eta$) pair which gave the lowest $\chi^2$ in each case.
These values are plotted in  Fig.\,\ref{first_cd}.
As in earlier work (Murgia et al. 2004; Guidetti et
al. 2008; Laing et al. 2008), we found a degeneracy between values of $B_{0}$ and
$\eta$, in the sense that the fitted values are positively correlated, but there
are clear minima in $\chi^2$ at both resolutions.  We therefore adopted the  
mean values of $B_0$ and $\eta$, weighted by $1/\chi^2$, as the best overall
estimates. These are also plotted in Fig.\,\ref{first_cd}
as blue crosses.
Although the central magnetic field strengths derived for the two RM images are
consistent at the 1$\sigma$ level ($B_{0}$=2.8$\pm$0.5\,$\mu$G and $B_{0}$=
4.1\,$\pm$1.2$\mu$G at 5.5 and 1.25-arcsec resolution, respectively), the
values of $\eta$ are not. The best-fitting values are $\eta$=0.0$\pm$0.1 at
5.5\,arcsec FWHM and  $\eta$=0.8$\pm$0.4 at
1.25\,arcsec FWHM. 

\begin{figure*}
\centering
\includegraphics[width=11cm]{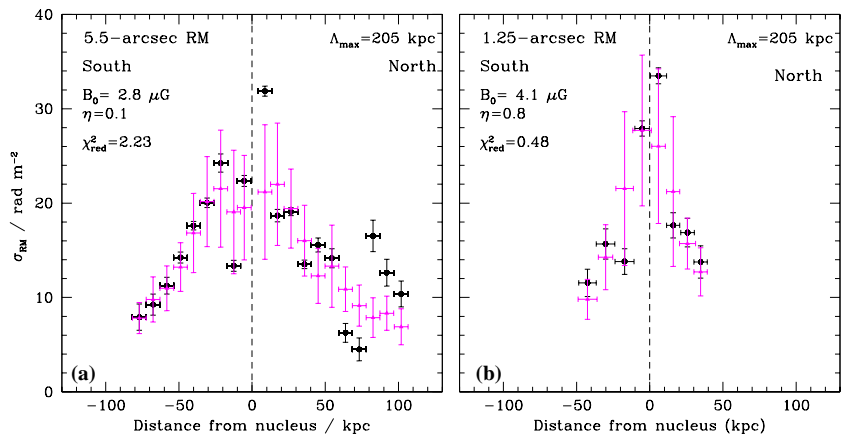}
\caption[]{(a): Observed and synthetic radial profiles  for rms Faraday \srm at
  5.5\,arcsec as functions of the projected distance from the radio source
  centre.  The outer scale is $\Lambda_{\rm max}$=205\,kpc.
  The black points represent the data with vertical bars
  corresponding to the rms error of the RM fit. The magenta triangles represent the
  mean values from 35 simulated profiles and the vertical bars are the rms
  scatter in these profiles due to sampling. (b) as (a) but
  at 1.25\,arcsec.\label{first}}
\end{figure*}

We next produced 35 RM simulations at each angular resolution by fixing  $B_0$ and $\eta$
at their best values for that resolution.  This allowed us to calculate 
weighted $\chi^2$'s for the \srm profiles, evaluating the errors for
each box by summing in quadrature the rms due to sampling (determined from the dispersion in
the realizations) and the fitting error of the observations. 
These values are listed in Table\,\ref{3dfit}.
The observed and best-fitting model profiles at both the angular resolutions
are shown in  Fig.\,\ref{first}.

A model with $\eta \simeq 0$ at all radii is unlikely a priori: 
previous work has found values of $0.5 \la \eta \la 1$ in other sources 
(Dolag 2006; Guidetti et al. 2008; Laing et al. 2008). The most likely
explanation for the low value of $\eta$ inferred from the low-resolution image
is that the electron density distribution is not well represented by a
$\beta$-model (which describes a spherical and smooth distribution) at large
radii. In support of this idea, Fig.\,\ref{XMM} shows that the morphology of the
X-ray emission is not spherical at large radii, but quite irregular.  Croston et
al. (2003) and Croston et al. (2008) pointed out that the quality of the fit of a single $\beta$-model to
the X-ray surface brightness profile was poor in the
outer regions, suggesting small-scale deviations in the gas distribution.  The
single $\beta$-model gave a better fit to the inner region of the X-ray surface
brightness profile, where the polarized emission of 3C\,449 can be observed at
1.25-arcsec resolution.
Our aim is to fit the \srm profiles at both resolutions with the same
distribution of
$n_{e}(r) \langle B(r)^2\rangle^{1/2}$.
In the rest of this subsection we assume that the density profile
$n_{e}(r)$ is still represented by the single $\beta$-model,
even though we have argued that it might not be appropriate
in the outer regions of the hot gas distribution.
Although the resulting estimates of field strength at large radii
may be unreliable, the fit is still necessary for the calculation of
outer scale described in Sec.\,\ref{sec:sfuncov}, which depends only on 
the combined spatial variation of density and field strength.

The best-fitting model at 1.25-arcsec resolution,
which is characterized by a more physically reliable $\eta$, gives a very bad
fit to the low resolution profile at almost all distances from the
core (Fig.\,\ref{first}a). Conversely, the model determined at 5.5-arcsec resolution  
gives a very poor fit to the sharp peak in \srm observed
within 20\,kpc of the nucleus at 1.25-arcsec resolution, where the radio and
X-ray data give the strongest constraints (Fig.\,\ref{first}b). We
have also verified that no single intermediate value of $\eta$ gives an adequate
fit to the \srm profiles at all distances from the nucleus.

A better description of the observed \srm profile is provided by the empirical
function:
\begin{eqnarray}
\langle B^2(r)\rangle^{1/2} &= & B_{0}\left[\frac{n_e(r)}{n_0}\right]^{~\eta_{int}} ~~~~~~~
r\leq{r_{\rm m}}         \nonumber \\
                           &= & B_{0} \left[\frac{n_e(r)}{n_0}\right]^{~\eta_{out}}  ~~~~~~
r>r_{\rm m}\,,~~~~~~~~~~~~~~~~~~~~~~~~~~~~~~~~\nonumber \\
\label{brb}
\end{eqnarray}
where $\eta_{\rm int}$, $\eta_{\rm out}$ are the inner and outer scaling index of the
magnetic field and  $r_{\rm m}$ is the break radius. 

We fixed $\eta_{\rm int}$=1.0 and $\eta_{\rm out}$=0.0, consistent with
our initial results, in order to reproduce both the inner sharp peak and the
outer flat decline of the \srm observed at the two resolutions, keeping $r_{\rm
m}$ as a free parameter. We made three sets of three-dimensional simulations
for values of the outer scale $\Lambda_{\rm max}$= 205, 65 and
20\,kpc. Anticipating the result of Sect.~\ref{sec:sfuncov}, we plot the results
only for $\Lambda_{\rm max}$= 65\,kpc, but the derived \srm profiles are in any
case almost independent of the value of the outer scale in this range.  The new simulations
were made only at a resolution of 5.5\,arcsec, since the larger field of view at
this resolution is essential to define the change in slope of the profile.

\begin{figure*}
\centering
\includegraphics[height=6cm]{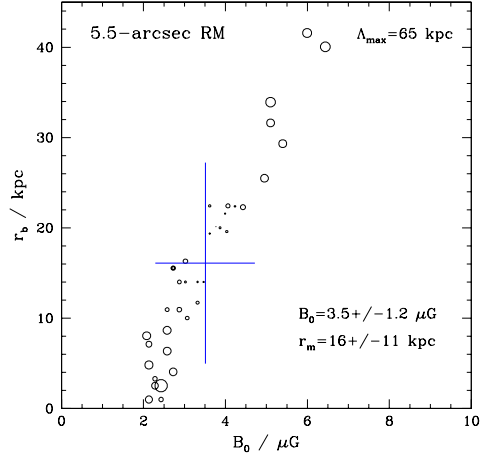}
\caption[]{Distribution of the best-fitting values of $B_{0}$ and $r_{\rm m}$ from 
35 sets of simulations, each covering ranges of 0.5 -- 10\,$\mu$G in $B_0$ and 0
-- 50\,kpc in $r_{\rm m}$. The sizes of the circles are proportional to $\chi^2$ and
the blue cross represents the means of the distribution
weighted by $1/\chi^2$.
The plot shows the degeneracy between $B_{0}$ and $r_{\rm m}$ described in the
text.}
\label{200_e}
\end{figure*}

In order to determine the best-fitting break radius, ${r_{\rm m}}$ in
Eq.\,\ref{brb}, we produced 35 sets of synthetic RM images for a grid of values
of $B_0$ and ${r_{\rm m}}$ for each outer scale, noting the pair of values which gave the minimum
unweighted $\chi^2$ for the \srm profile for each set of simulations.  These
values are plotted in Fig.~\ref{200_e}, which shows that there is a degeneracy
between the break radius $r_{\rm m}$ and $B_0$.  As with the similar degeneracy
between $B_0$ and $\eta$ noted earlier, there is a clear minimum in $\chi^2$,
and we therefore adopted the mean values of $B_0$ and $r_{\rm m}$ weighted by
$1/\chi^2$ as our best estimates of the magnetic-field parameters.

We then made 35 simulations with the best-fitting values of $B_0$ and
$\eta$ for each outer scale and evaluated the weighted $\chi^2$'s for the
resulting \srm profiles.  All three values of $\Lambda_{\rm max}$ we
investigated give reasonable fits to the observed \srm profile along the whole
radio source. The fit for $\Lambda_{\rm max} = 65$\,kpc is marginally better
than for the other two values ($\chi^2_{\rm red} = 1.8$), consistent with the results of
Sect.~\ref{sec:sfuncov}, below. In
this case the central magnetic field strength is 3.5$\pm1.2$\,$\mu$G and the
break radius is 16$\pm$11\,kpc. For the power spectrum with $\Lambda_{\rm max}
= 65$\,kpc and these best-fitting parameters, we also produced three-dimensional
simulations at a resolution of 1.25\,arcsec.  Even though the fitting procedure
is based only on the low-resolution data, this model also reproduces the
1.25-arcsec profile very well ($\chi^2_{\rm red}$=0.7).  Combining the values of
$\chi^2$ for the two resolutions, using the 1.25-arcsec profile close to the
core and the 5.5-profile at larger distances, we find $\chi^2_{\rm red}$=1.8.

\begin{figure*}
\centering
\includegraphics[height=10.5cm]{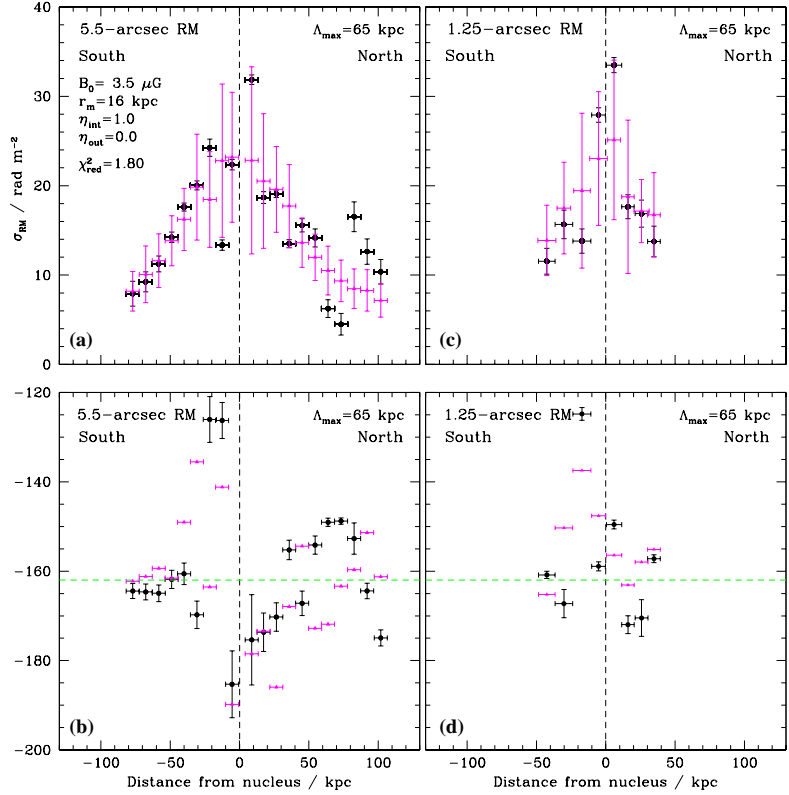}
\caption[]{Comparison between observed and synthetic profiles of rms and mean
Faraday rotation at resolutions of 5.5\,arcsec (a and b) and 1.25\,arcsec (c and
d).  The synthetic profiles are derived from the best-fitting model with
$\Lambda_{\rm max} = 65$\,kpc.  The black points represent the data with
vertical bars corresponding to the rms fitting error. 
(a) and (c): Profiles of \srm. The magenta
triangles represent the mean values from 35 simulations at 5.5-arcsec resolution
and the vertical bars are the rms scatter in these profiles due to sampling.
(b) and (d): Profiles of \rmm derived from single example realizations at 5.5 and 1.25-arcsec resolution.
}
\label{200}
\end{figure*}

Fig.\,\ref{200} shows a comparison of the observed radial profiles
for rms Faraday \srm  and \rmm  with the synthetic ones
derived for this model. 
The synthetic \srm profile plotted in Fig.~\ref{200} is the mean over 35 simulations, and may
be compared directly with the observations.
In contrast, the \rmm profile is derived from a single example realization.
It is important to emphasize that the latter is {\em one example of a random process}, and is 
not expected to fit the observations; rather, we aim to compare the fluctuation
amplitude as a function of position.

The values of $B_0$ and $\chi^2_{\rm red}$  for all
of the three-dimensional simulations, together with $\eta$ and $r_{\rm m}$ for
the single and double power-law profiles, respectively, are summarized in
Table\,\ref{3dfit}.

\begin{table*} 
\centering
\caption{Results from the three-dimensional fits at both the angular resolutions of 1.25 and 5.5\,arcsec.\label{3dfit}}     
\begin{tabular}{c c c c c c c c}     
\hline\hline       
  FWHM    & $\Lambda_{\rm max}$     & \multicolumn{3}{c} {single $\eta$} & \multicolumn{3}{c} {broken $\eta$}  \\
     (arcsec)    &   (kpc)           &   $B_{0}$ ($\mu$G) & $\eta$ & $\chi^2_{\rm red}$  & $B_{0}$ ($\mu$G) & $r_{\rm m}$(kpc) & $\chi^2_{\rm red}$\\
\hline      
1.25    &  205                & 4.1$\pm$1.1    &  0.8$\pm$0.4               &  0.48    & - & - & - \\
5.50    &  205                & 2.8$\pm$0.5     &  0.1$\pm$0.1         &  2.2 & 3.5$\pm$0.7  & 17$\pm$9 & 1.9 \\
5.50    &  65                 & -        & -                  & -   & 3.5$\pm$1.2 & 16$\pm$11  & 1.8 \\
5.50    &  20                 & -   &     -                   & -   & 3.5$\pm$0.8 & 11$\pm$8  & 2.1 \\
\hline
\end{tabular}
\end{table*}

\begin{figure*}
\centering
\includegraphics[width=9cm]{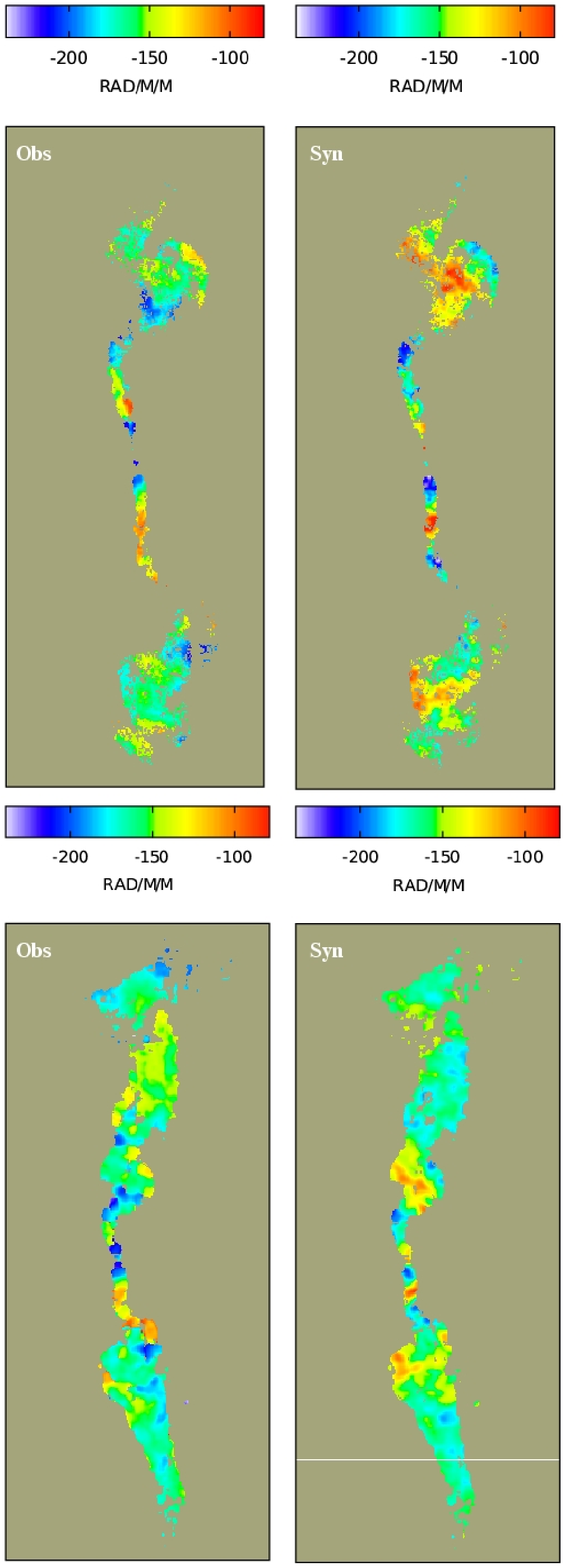}
\caption[]{Comparison of observed and representative synthetic distribution of Faraday RM at 1.25\,arcsec and 5.5\,arcsec.
The synthetic images have been produced for the best-fitting model with $\Lambda_{\rm max}$=65\,kpc.
The colour scale is the same for all displays.
\label{comp}}
\end{figure*}

\subsection{The outer scale of magnetic-field fluctuations}
\label{sec:sfuncov}

The theoretical RM structure function for uniform field strength, density and
path length and a power spectrum with a low-frequency cut-off should
asymptotically approach a constant value ($2\sigma^2_{\rm RM}$  for a large enough averaging
region) at separations $\ga \Lambda_{\rm max}$. 
The observed RM structure function
of the whole source is heavily modified from this theoretical one by the scaling
of the electron gas density and magnetic field at large separations, which acts 
to suppress power on large spatial scales.
In Sect.\,\ref{2d} we therefore
limited the study of the structure function to sub-regions of 3C\,449 in which
uniformity of field strength, density and path length (and therefore of the
power-spectrum amplitude) is a reasonable assumption, inevitably limiting our
ability to constrain the power spectrum on the largest scales. 

Now that we have an adequate model for the variation of $n_e(r)\langle B^2(r)
\rangle^{1/2}$ with radius (Eq.~\ref{brb}), we can correct for it to derive 
what we call the pseudo-structure function -- that is the structure function for
a power-spectrum amplitude which is constant over the source. This can be
compared directly with the structure functions derived from the Hankel transform
of the power spectrum. To evaluate the pseudo-structure function, we
divided the observed 5.5-arcsec RM image by the function:
\begin{equation}
\label{detrend}
\left [\int_{0}^{L}n_{e}(r)^2B(r)^2dl\right ]^{1/2}
\end{equation}
(En{\ss}lin \& Vogt 2003), where the radial variations of $n_e$ and B are those of the best-fitting model
(Sect.\,\ref{sec:model}) and the upper integration limit $L$ has a length of 10 times the core radius.
The integral was normalized to unity at
the position of the radio core: this is equivalent to fixing the field and gas
density at their maximum values and holding them constant everywhere in the
group. The normalization of the pseudo-structure function should then be quite
close to that for the two central jet regions.

\begin{figure*}
\centering
\includegraphics[width=17cm]{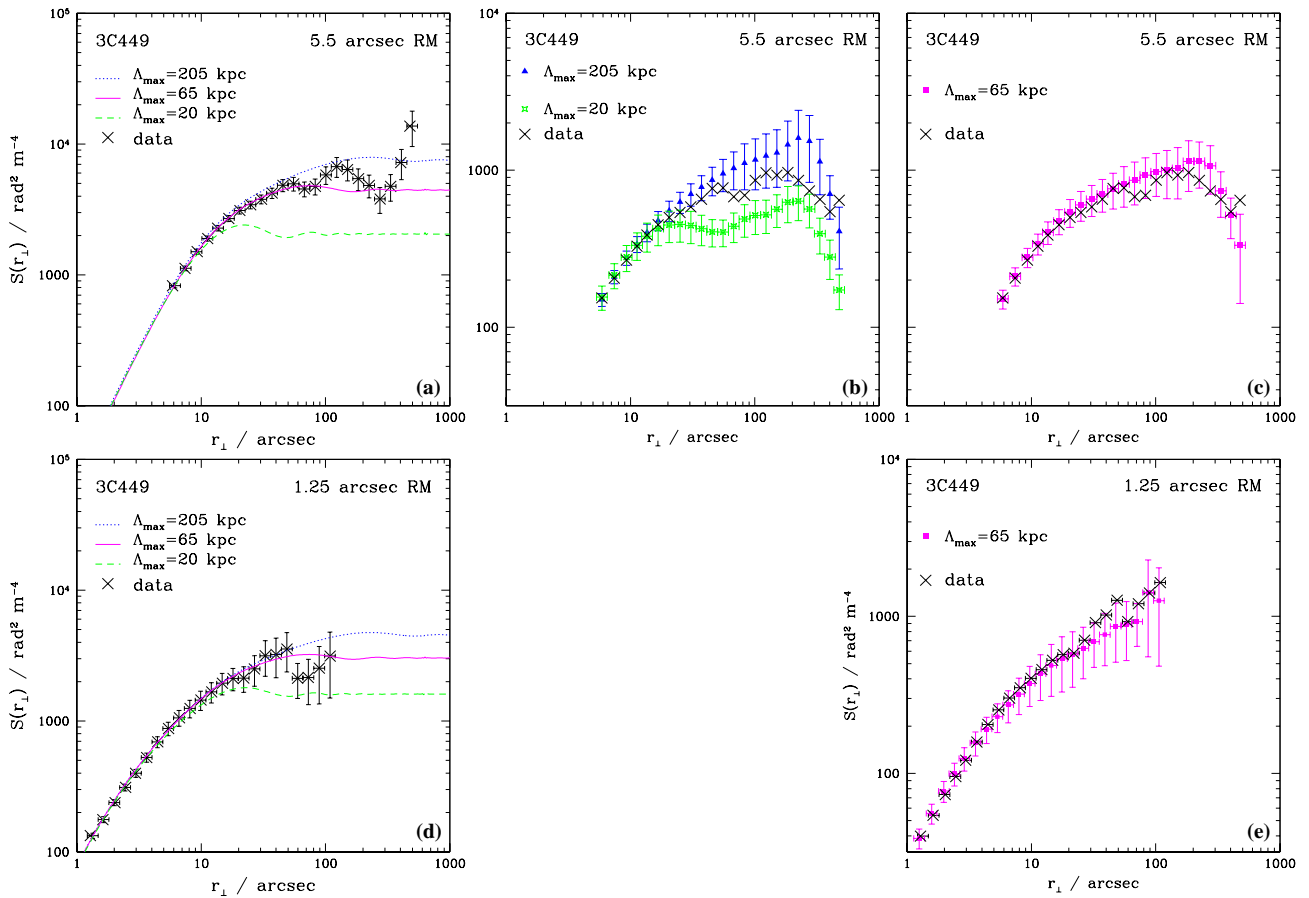}
\caption[]{(a): Comparison of the observed pseudo-structure function of the
5.5-arcsec RM image as described in the text (points) with the predictions of the BPL model (curves, derived using a Hankel transform). The predicted curves are for
$\Lambda_{\rm max}$=205 (blue dotted), 65 (continuous magenta) and 20\,kpc (green dashed).
(b) and (c): structure functions of the
observed (points) and synthetic 5.5-arcsec RM images produced with the $\Lambda_{\rm max}$=205,20\,kpc
and $\Lambda_{\rm max}$=65\,kpc, respectively.
(d) and (e) as (a) and (c) respectively, but at 1.25-arcsec resolution.
The error bars are the rms from 35 structure functions at a given $\Lambda_{\rm max}$, 
in (a) and (d) are representative error bars from the model with  $\Lambda_{\rm max}$=65\,kpc. 
}
\label{sfuncov}
\end{figure*}

The pseudo-structure function is shown in Fig.\,\ref{sfuncov}(a) together
with the predictions for the BPL power spectra with $\Lambda_{\rm max}$=205, 65 and 20\,kpc.
As expected, the normalization of the pseudo-structure function is consistent
with that of the jets (Fig.\,\ref{sfunc}a and b).

The comparison between the synthetic and observed pseudo-structure functions
indicates firstly that they agree very well at small separations, independent of
the value of $\Lambda_{\rm max}$.  This confirms that the best BPL power
spectrum found from a combined fit to all six sub-regions is a very good fit
over 
the entire source.  Secondly, despite the poor sampling on  very large scales, 
the asymptotic values of the predicted structure functions for the three values
of $\Lambda_{\rm max}$ are sufficiently different from each other that we can
determine an approximate outer scale.  The model with
$\Lambda_{\rm max}$=65\,kpc gives the best representation of the data. The fit
is within the estimated errors except for a marginal discrepancy at the very
largest (and therefore poorly sampled) separations. That 
with $\Lambda_{\rm max}$=20\,kpc is inconsistent with the observed
pseudo-structure function for any separation $\ga$20\,arcsec ($\simeq$6\,kpc) where
the sampling is still very good, and is firmly excluded.  The model with
$\Lambda_{\rm max}$=205\,kpc has slightly, but significantly too much power on
large scales. 

  We emphasize that our estimate of the outer scale of the RM
  fluctuations is essentially independent of the functional form assumed for the
  variation of the field strength with radius in the central region, which
  affects the structure function only for small separations. Our results are
  almost identical if we fit the field-strength variation with 
  either the profile of Eq.\,\ref{br} (with $\eta \approx 0$) or that of Eq.\,\ref{brb}.  

As for the structure functions of individual regions, the pseudo-structure
function at large separations is clearly affected by poor sampling: this
increases the errors but does not produce any bias in the derived 
values. At large radii, however, the integral in equation~\ref{detrend} becomes
small, so the noise on the RM image is amplified. This is a potential source of
error, and we have therefore checked our
results using numerical simulations.
We calculated the mean and rms structure functions for sets of realizations of
RM images generated using the FARADAY code, as
in Sect.~\ref{3Dcode}, for different values of  $\Lambda_{\rm max}$.
These structure functions are plotted in Figs.\,\ref{sfuncov}(b) and (c).

The main difference between the model structure functions derived from
simulations and the pseudo-structure functions described earlier is that the
former show a steep decline in power on large scales in place of a plateau. This
occurs because the smooth fall-off in density and magnetic field strength with
distance from the nucleus suppresses the fluctuations in RM on large scales.

The results of the simulations confirm our analysis using the pseudo-structure
function.  The mean model structure function with $\Lambda_{\rm max} = 65$\,kpc
again fits the data very well, except for a marginal discrepancy at the largest
scales. In view of the deviations from spherical symmetry on large
scales evident in the X-ray emission surrounding 3C\,449 (Fig.~\ref{XMM}), we do
not regard this as a significant effect.

In order to check that the best-fitting density and field model also 
reproduces the data at small separations, we repeated the analysis at 
1.25-arcsec resolution. The observed pseudo-structure function is shown in 
Fig. 14(d), together with the the predictions for the BPL power spectrum 
with $\Lambda_{\rm max}$= 205, 65 and 20 kpc. The observed structure function is 
compared with the mean from 35 simulations with $\Lambda_{\rm max}$= 65 kpc in 
Fig. 14(e). In both cases, the agreement for $\Lambda_{\rm max}$= 65 kpc is 
excellent.

In Fig.\,\ref{comp}, example realizations of this model with the best-fitting
field variation are shown for resolutions of 1.25 and 5.5\,arcsec alongside the
observed RM images. 

\section{Summary and comparison with other sources}
\label{sec:sum}

\subsection{Summary}

In this work we have studied the structure of the magnetic field associated with
the ionized medium around the radio galaxy 3C\,449.  We have analysed images of
linearly polarized emission with resolutions of 1.25\,arcsec and 5.5\,arcsec
FHWM at seven frequencies between 1.365 and 8.385 GHz, and produced images of
degree of polarization and rotation measure.  The RM images at both the angular
resolutions show patchy and random structures. In order to study the spatial
statistics of the magnetic field, we used a structure-function analysis and
performed two- and three-dimensional RM simulations.  We can summarize the
results as follows.
\begin{enumerate}
\item The absence of deviations from $\lambda^2$ rotation over a wide range of
  polarization position angle implies that a pure foreground Faraday screen with no mixing of radio-emitting and
  thermal electrons is a good approximation for 3C\,449
  (Sect.\,\ref{sec:rm_obs}).
\item The dependence of the degree of polarization on wavelength
  is well fitted by a Burn law. This is also consistent with pure foreground
  rotation, with the residual depolarization observed at the higher resolution
  being due to unresolved RM fluctuations across the beam
  (Sect.~\ref{sec:dp}). There is no evidence for detailed correlation of
   radio-source structure with either RM or depolarization.
 \item There is no obvious anisotropy in the RM distribution, consistent with our 
   assumption that the magnetic field is an isotropic, Gaussian random variable.
 \item Our best estimate for the Galactic contribution to the RM of 3C\,449 is a
   constant value of  $-$160.7\,rad\,m$^{-2}$ (Sect.\,\ref{sec:gal}).
 \item The RM structure functions for six different regions of the source are
   consistent with the hypothesis that only the amplitude of the RM power
   spectrum varies across the source.
 \item A broken power-law spectrum of
 the form given in Eq.\,\ref{bpl}  with  $q_{\rm{l}}$= 2.07,
 $q_{\rm{h}}$=2.98, $f_{\rm b}$=0.031\,$\rm{arcsec^{-1}}$ and $f_{\rm max} =
 1.67$\,arcsec$^{-1}$ (corresponding to a spatial scale $\Lambda_{\rm
 min}$=0.2\,kpc) 
 is consistent with the observed structure functions and
 depolarizations for all six regions.  No single power law provides a good fit
 to all of the structure functions.
\item The high-frequency cut-off in the power spectrum is required to model the
  depolarization data.
 \item The profiles of \srm strongly suggest that most of the fluctuating component of RM
   is associated with the intra-group gas, whose core radius is comparable with
   the characteristic scale of the profile (Sect.~\ref{sec:rm_images}). The
   symmetry of the profile is consistent with the idea that the radio source axis
   is close to the plane of the sky.
\item We therefore simulated the RM distributions expected for an isotropic,
  random magnetic field in the hot plasma surrounding 3C\,449, assuming the
  density model derived by Croston et al.\ (2008).
\item These three-dimensional simulations show that the dependence of magnetic
  field on density is best modelled by a broken power-law function with $B(r)
  \propto n_e(r)$ close to the nucleus and $B(r) \approx$ constant at larger
  distances.
\item With this density model, our best estimate of the central magnetic field
  strength is $B_0 = 3.5 \pm 1.2 \mu$G. 
 \item Assuming these variations of density and field strength with radius, a
 structure-function analysis can be used to estimate the outer scale $\Lambda_{\rm max}$ of the
 magnetic-field fluctuations. We find excellent agreement for $\Lambda_{\rm max}
 \approx 65$\,kpc ($f_{\rm min} =
0.0053$\,arcsec$^{-1}$).
\end{enumerate} 

\subsection{Comparison with other sources}

Our results are qualitatively similar to those of Laing et al.\ (2008) on
3C\,31. The maximum RM fluctuation amplitudes are similar in the two sources, as
are their environments. For spherically-symmetric gas density models, the
central magnetic fields are almost the same: $B_0 \approx 2.8\mu$G for 3C\,31
and $3.5\mu$G for 3C\,449.  Both results are consistent with the idea that the
RM fluctuation amplitude in galaxy groups and clusters scales roughly linearly
with density, ranging from a few\,rad\,m$^{-2}$ in the much sparsest
environments (e.g.\ NGC\,315; Laing et al. 2006), through intermediate values 
$\approx$~30 -- 100\,rad\,m$^{-2}$ in rich groups such as 3C\,31 and 3C\,449 to $\sim
10^4$\,rad\,m$^{-2}$ in the centres of clusters with cool cores.

The RM distribution of 3C\,31 is asymmetrical, the
northern (approaching) side of the source showing a much lower fluctuation
amplitude, consistent with the inclination of $\approx 50^\circ$ estimated by
Laing \& Bridle (2002). Detailed modelling of the RM profile led Laing et al.\
(2008) to suggest that there is a cavity in the X-ray gas, but this would have
to be significantly larger than the observed extent of the radio lobes and is,
as yet, undetected in X-ray observations. A broken power-law scaling of magnetic
field with density, similar to that found for 3C\,449 in the present paper,
would also improve the fit to the \srm profile for 3C\,31; alternatively, the
effects of cavities around the inner lobes and spurs of 3C\,449 might be
significant. Deeper X-ray observations of both sources are needed to resolve
this issue.  In neither case is the magnetic field dynamically important: for
3C\,449 we find that the ratio of the thermal and magnetic-field pressures is
$\approx$30 at the nucleus
and $\approx$400 at the core radius of the group gas, $r_c = 19$\,kpc. The
magnetic field is therefore not dynamically important, as in 3C\,31.

The magnetic-field power spectrum in both sources can be fit by a broken
power-law form.  The low-frequency slopes are 2.1 and 2.3 for 3C\,449 and 3C\,31
respectively. In both cases, the power spectrum steepens at higher spatial
frequencies, but for 3C\,31 a Kolmogorov index (11/3) provides a good fit,
whereas we find that the depolarization data for 3C\,449 require a cut-off below
a scale of 0.2\,kpc and a high-frequency slope of 3.0. The break scales are
$\approx$5\,kpc for 3C\,31 and $\approx$11\,kpc for 3C\,449.  It is important to
note that the simple parametrized form of the power spectrum is not unique, and
that a smoothly curved function would fit the data equally well.

The gas-density structure on large scales in the 3C\,31 group is uncertain, so
Laing et al.\ (2008) could only give a rough lower limit to the outer scale of
magnetic-field fluctuations, $\Lambda_{\rm max} \ga 70$\,kpc.  For 3C\,449, we
find $\Lambda_{\rm max} \approx 65$\,kpc.  The projected distance between 3C\,449 and
its nearest neighbour is $\approx$33\,kpc (Birkinshaw et al. 1981), similar to
the scale on which the jets first bend through large angles (Fig.~\ref{XMM}). As
in 3C\,31, it is plausible that the outer scale of magnetic-field fluctuations
is set by interactions with companion galaxies in the group.

\begin{acknowledgements}
This work is part of the ``Cybersar'' Project, which is managed by the 
COSMOLAB Regional Consortium with the financial support of the Italian 
Ministry of University and Research (MUR), in the context of the ``Piano 
Operativo Nazionale Ricerca Scientifica, Sviluppo Tecnologico, Alta 
Formazione (PON 2000-2006)''.
We thank Luigina Feretti for providing the 
VLA data of 3C449, Greg Taylor for the use of his rotation measure code
and Marco Bondi for many
helpful comments.
We also acknowledge the use of HEALPIX package (http://healpix.jpl.nasa.gov)
and the provision of the models of Dineen \& Coles (2005) in HEALPIX format.
\end{acknowledgements}

\end{document}